\documentclass[aps,a4paper, preprint, superscriptaddress,preprintnumbers,floatfix,nofootinbib,amsmath,amssymb]{revtex4}
\usepackage{url}
\usepackage{hyperref}
\usepackage{cleveref}

\usepackage{amstext,amssymb}
\usepackage{amsmath}
\usepackage{graphicx}
\usepackage{xspace}
\usepackage{color}
\usepackage{units}
\usepackage{array}
\usepackage{feynmp}
\usepackage{amsmath,bm}
\usepackage{amssymb}
\usepackage{textcomp}
\usepackage{slashed} % feynman slash notation via \slashed{p}
\usepackage{multirow}
\newcommand{\be}{\begin{equation}}
\newcommand{\ee}{\end{equation}}
\newcommand{\bea}{\begin{eqnarray}}
\newcommand{\eea}{\end{eqnarray}}
\newcommand{\nn}{\nonumber}
\usepackage{epstopdf}
\usepackage{float}
\usepackage{multirow}

\def\s1{\hat s}

\usepackage{slashed}

\newcommand{\PreserveBackslash}[1]{\let\temp=\\#1\let\\=\temp}
\newcolumntype{C}[1]{>{\PreserveBackslash\centering}p{#1}}
\newcolumntype{R}[1]{>{\PreserveBackslash\raggedleft}p{#1}}
\newcolumntype{L}[1]{>{\PreserveBackslash\raggedright}p{#1}}
\newcommand{\nua}[1]{\ensuremath{\rlap{\kern-2.5pt\ensuremath{\overset{\scriptscriptstyle(-)}{\phantom{\nu}}}}{\ensuremath{{\nu}_{#1}}}}\xspace}

\begin{document}
\title{ Inverse seesaw in $A_5^\prime$ modular symmetry}
\author{ Mitesh Kumar Behera}
\email{miteshbehera1304@gmail.com}
\affiliation{School of Physics,  University of Hyderabad, Hyderabad - 500046,  India}

\author{ Rukmani Mohanta}
\email{rmsp@uohyd.ac.in}
\affiliation{School of Physics,  University of Hyderabad, Hyderabad - 500046,  India}

\begin{abstract}
We make an investigation of modular $\Gamma^{\prime}_5 \simeq A^{\prime}_5$ group in inverse seesaw framework.  Modular symmetry is advantageous because it reduces the usage of extra scalar fields significantly. Moreover, the Yukawa couplings are expressed in terms of Dedekind eta functions, which also have a $q$ expansion form, utilized to achieve numerical simplicity. Our proposed model includes six heavy fermion superfields i.e., $\mathcal{N}_{Ri}$, $\mathcal{S}_{Li}$ and a weighton.  The study of neutrino phenomenology becomes simplified and effective by the usage of $A^\prime_5$ modular symmetry, which provides us a well defined mass structure for the lepton sector. Here, we observe that all the neutrino oscillation parameters, as well as  the effective electron neutrino mass in neutrinoless  double beta decay can be accommodated   in this model.  We also briefly discuss  the  lepton flavor violating decays $\ell_i \to \ell_j \gamma$  and comment on non-unitarity of lepton mixing matrix.
\end{abstract}

\maketitle
\flushbottom

\section{INTRODUCTION}
\label{sec:intro}

The results from various neutrino oscillation experiments have unambiguously established the fact that neutrinos posses very small but non-zero masses contradicting their vanishing mass concept presumed in the  Standard Model (SM).  Therefore, understanding the origin of the neutrino mass necessitates to employ  physics beyond the SM.   One of the conventional ways to generate the light neutrino masses is through the canonical  seesaw mechanism \cite{Mohapatra:1979ia,Brdar:2019iem,Branco:2019avf,Bilenky:2010zza}, where three heavy right handed (RH) neutrinos $\mathcal{N}_{R_i}$ are introduced on top of the  SM particle spectrum. The inclusion of right-handed neutrinos  not only generates the Dirac mass term but also leads to Majorana mass for $\mathcal{N}_{Ri}$'s,  of the form $\overline{\mathcal{N}_{Ri}} \mathcal{N}_{Ri}^c$ which violates the lepton number by two units. The master formula  for generating  the masses of the active neutrinos is governed by $m_\nu \approx -\mathcal{M}_D \mathcal{M}^{-1}_R \mathcal{M}^T_D$, where $\mathcal{M}_D$  is the Dirac neutrino mass matrix and $\mathcal{M}_{R}$ being the Majorana neutrino mass matrix of the heavy RH neutrinos,  satisfying the relation $\mathcal{M}_D$ $\ll \mathcal{M}_{R}$.  However, myriad literature on seesaw models show work on other extensions like type-II, with the inclusion of a scalar triplet \cite{Gu:2006wj,Luo:2007mq,Antusch:2004xy,Rodejohann:2004cg,Gu:2019ogb,McDonald:2007ka}, type-III \cite{Liao:2009nq,Ma:1998dn,Foot:1988aq,Dorsner:2006fx,Franceschini:2008pz,He:2009tf},  where a fermion triplet is added to the SM particle content. 
In these approaches, the masses of the new heavy particles are quite heavy and are beyond the access of the present and future experiments.

Many other alternative approaches were proposed, e.g., linear seesaw \cite{Ma:2009du,Gu:2010xc,Sruthilaya:2017mzt,Borah:2018nvu}, inverse seesaw \cite{Das:2012ze,Arganda:2014dta,Ma:2015raa,Dias:2012xp,Dev:2012sg,Dias:2011sq,Bazzocchi:2010dt},  where the new physics scale responsible for neutrino mass generation can be brought down to  TeV scale, at the expense  of the inclusion  of new additional  fermion fields ($S_i$), which are SM singlets.  
 The inverse seesaw formalism is implemented by including three additional left handed (LH) singlet fermions $\mathcal{S}_{L_i}$ and hence, the basis that involves for  the neutrino mass generation is  $(\nu_L^c, \mathcal{N}_{R_i}, \mathcal{S}_{L_i})^T$. This leads to the neutrino mass matrix structure as $m_\nu \approx (\mathcal{M}_D/ \mathcal{M}_{RS}) ~\mathcal{M}_\mu~( \mathcal{M}_D/\mathcal{M}_{RS})^T$, where  $\mathcal{M}_\mu$ is the Majorana mass term for the heavy singlet fermion $S_{Li}$.    For  inverse seesaw, the various mass terms satisfy the relation   $\mathcal{M}_\mu \ll \mathcal{M}_D< \mathcal{M}_{RS}$, and hence, the neutrino mass is given by $m_\nu \approx$ $ \mathcal{M}_D^2\mathcal{M}_\mu/\mathcal{M}_{RS}^2$.  So to get the correct   order of the light neutrino masses, the typical values of different mass scales are:  $\mathcal{M}_D \sim $ 10 GeV, $\mathcal{M}_{RS} \sim 10~ \rm{TeV}$, and $\mathcal{M}_\mu \sim 1~ \rm{keV}$.
 
 Genearally, to implement inverse seesaw certain symmtries are assumed, like discrete flavour symmetries  $S_3$ \cite{CarcamoHernandez:2013krw,Ma:2014qra}, $A_4$ \cite{CarcamoHernandez:2017kra,Borah:2017dmk,Hirsch:2009mx,Kalita:2015jaa,Altarelli:2010gt}, $S_4$ \cite{Ma:2005pd,Dorame:2012zv,CarcamoHernandez:2019eme} etc., to avoid certain unwanted terms in the extended neutrino mass matrix of $(\nu_L^c, \mathcal{N}_{R_i}, \mathcal{S}_{L_i})^T$ basis. However, a number of flavon fields are required for the breaking of these flavor symmetries as well as to accommodate the observed neutrino oscillation data and the vacuum alignment of these flavon fields pose a challenging task.  
But in recent times,  modular symmetry \cite{Kobayashi:2018vbk,Feruglio:2017spp,deAdelhartToorop:2011re} has gained pace and is in the limelight. Modular symmetry removes the usage of excess flavon fields, where, the role of flavons is performed by Yukawa couplings, which are holomorphic function of modulus $\tau$. When this modulus acquires the vacuum expectation value (VEV), it breaks the flavor symmetry. Exploration of myriad text shows work on modular groups $S_3$ \cite{Du:2020ylx,Mishra:2020gxg,Okada:2019xqk}, $S_4$ \cite{Penedo:2018nmg,Novichkov:2018ovf,Okada:2019lzv}, $A_4$ \cite{Abbas:2020vuy,Nagao:2020snm,Asaka:2020tmo,Nomura:2020opk,Okada:2020dmb,Behera:2020lpd,Behera:2020sfe,Ding:2019zxk,Altarelli:2005yx},  $A_5$ \cite{Novichkov:2018nkm,Yao:2020zml}, double covering of $A_4$ \cite{Liu:2019khw}, double covering of $A_5$ \cite{Wang:2020lxk}. These modular groups help to accurately calculate the neutrino oscillation parameters  at $3\sigma$ level along with other observables.

In this work,  we intend to  focus on the  double covering modular group  $\Gamma^{\prime}_5 \simeq A^{\prime}_5$ and its implications on neutrino phenomenology. In the past,  quite a few works in the literature have been discussed  the significace of finite groups, which comprehend the basic properties of  $A^\prime_5$ group ~\cite{Everett:2010rd,Hashimoto:2011tn,Chen:2011dn}.  So here, we mention only the essential points regarding $A^\prime_5$ modular symmetry group. The $A^\prime_{5}$ group has 120 elements, which can be constructed by three generators $S$, $T$ and $R$, which satisy the identities $S^2=R$, $(ST)^3= \mathbb{I}$, $R^2= \mathbb{I}$ and $RT=TR$ \cite{Wang:2020lxk}. These 120 elements are categorized into nine conjugacy classes, which classifies them as the nine distinct irreducible representations, symbolized as ${\bf 1}$, $\widehat{\bf 2}$, $\widehat{\bf 2}^{\prime}_{}$, ${\bf 3}$, ${\bf 3}^{\prime}_{}$, ${\bf 4}$, $\widehat{\bf 4}$, ${\bf 5}$ and $\widehat{\bf 6}$ by their dimensions. Moreover, conjugacy classes and character table of $A^\prime_5$, as well as the representation matrices of all three generators $S$, $T$ and $R$ in the irreducible representations, are presented in Appendix~\cite{Wang:2020lxk}.  It should be noted  that the  ${\bf 1}$, ${\bf 3}$, ${\bf 3}^{\prime}_{}$, ${\bf 4}$ and ${\bf 5}$ representations with $R = \mathbb{I}$ coincide with those for $A^{}_{5}$, whereas $\widehat{\bf 2}$, $\widehat{\bf 2}^{\prime}_{}$, $\widehat{\bf 4}$ and $\widehat{\bf 6}$ are unique for $A^\prime_5$ with $R = -\mathbb{I}$. As we are working in the modular space of $\Gamma(5)$, hence, its dimension is $5k+1$, where, $k$ is the modular weight.  A brief discussion concerning the modular space of $\Gamma(5)$ is presented in Appendix \ref{mod_space}. For $k=1$, the modular space $M_1[\Gamma(5)]$ will have six basis vectors i.e ($\widehat{e}_i$, where $i=1,2,3,4,5,6$) whose $q$-expansion is given below and they are used in expressing the Yukawa coupling $Y^{(1)}_{\widehat{\bf6}}$ as shown in Appendix \ref{HOYC}: 
\begin{eqnarray}
\widehat{e}^{}_1 & = & 1 + 3q + 4q^2_{} + 2q^3_{} + q^4_{} + 3q^5_{} + 6q^6_{} + 4q^7_{} - q^9_{} + \cdots \; , \nonumber \\
\widehat{e}^{}_2 & = & q^{1/5}_{} \left( 1 + 2q + 2q^2_{} + q^3_{} + 2q^4_{} + 2q^5_{} + 2q^6_{} + q^7_{} + 2q^8_{} + 2q^9_{} + \cdots \right) \nonumber \; , \\
\widehat{e}^{}_3 & = & q^{2/5}_{} \left( 1 + q + q^2_{} + q^3_{} + 2q^4_{} + q^6_{} + q^7_{} + 2q^8_{} + q^9_{} + \cdots \right) \; ,\nonumber \\
\widehat{e}^{}_4 & = & q^{3/5}_{} \left( 1 + q^2_{} + q^3_{} + q^4_{} - q^5_{} + 2q^6_{} + 2q^8_{} + q^9_{} + \cdots \right) \; ,  \nonumber \\
\widehat{e}^{}_5 & = & q^{4/5}_{} \left( 1 - q + 2q^2_{} + 2q^6_{} - 2q^7_{} + 2q^8_{} + q^9_{} +\cdots \right) \; , \nonumber \\
\widehat{e}^{}_6 & = & q \left( 1 - 2q + 4q^2_{} - 3q^3_{} + q^4_{} + 2q^5_{} - 2q^6_{} + 3q^8_{} - 2q^9_{} + \cdots \right) \; .
\label{eq:basisq}
\end{eqnarray} 

Structure of this paper is as follows. In Sec. \ref{sec:inverse}, we discuss the model framework for generating the light neutrino masses  using inverse seesaw mechanism with discrete $A^\prime_5$ modular flavor symmetry. This $A^\prime_5$ modular symmetry is double covered hence, there are more number of  irreducible representation as compared to $A_5$ modular symmetry. This helps us to construct charged leptons and neutral lepton mass matrices.   In Sec. \ref{sec:results}, numerical correlational study between the observables of neutrino sector and the model input parameters is established. A brief  discussion on the non-unitarity effect is presented in Sec. \ref{sec:non-unitarity}. In addition, lepton flavor violation (LFV) in the context of the present model is presented in Sec. \ref{sec:LFV} and  in Sec. \ref{sec:con},  we conclude our results.

\section{MODEL FRAMEWORK}
\label{sec:inverse}
We consider a scenario in which inverse seesaw is implemented in the context of supersymmetry (SUSY) to study the neutrino phenomenology, where the SM is extended with a discrete $A^\prime_5$ modular symmetry. An additional  local $U(1)_{B-L}$ symmetry is added to prohibit certain undesirable terms in the superpotential. The SM particle spectrum is supplemented with three extra RH singlet fermion superfields ($\mathcal{N}_{R_i}$), three LH singlet fermion superfields ($\mathcal{S}_{L_i}$) and one weighton ($\zeta$). The added fermion superfields of the model transform as $3^\prime$ under the $A^\prime_5$ modular group, whereas, the $U(1)_{B-L}$ charges assigned to them are $-1$ ($\mathcal{N}_{R_i}$) and 0 ($\mathcal{S}_{L_i}$). Also RH neutrinos are assigned modular weight 6 and LH neutrinos with 0.  The particle content and their charges under various groups are provided in Table \ref{tab:fields-inverse}.  
The $A^\prime_5$ and $U(1)_{B-L}$ symmetries are considered to be broken at a scale much higher than the electroweak symmetry breaking \cite{Dawson:2017ksx}. The $U(1)_{B-L}$ symmetry is spontaneously 
broken  by assigning non-zero vacuum expectation value (VEV) to the singlet weighton $\zeta$, and consequently  the additional singlet fermion superfields acquire their masses. 
 In addition to above, several higher order Yukawa couplings are introduced which obey the rule: $k_{Y} = k_{I^{}_1} + k_{I^{}_2} + \cdots + k_{I^{}_n}$, where $k^{}_{Y}$ is the weight on the Yukawa couplings and $k_{I^{}_i} (i=1,2,3,4\cdots)$ are the weights on the superfields.  These higher order Yukawa couplings implicitly depend on $Y^{(1)}_{\widehat{\bf 6}}$ whose complete forms are shown in Appendix \ref{HOYC}.
%%%%%%%%%%%%%%%%%%%%%%%%%%%%%%%%
\begin{center} 
\begin{table}[h!]%[tbh!]
%\begin{tiny}
\centering
\begin{tabular}{|c||c|c|c|c|c|c||c|c|c|c|}\hline\hline  
% & \multicolumn{8}{c||}{Fields}   \\ \hline \hline

\textbf{Fields} & ~$e_R$~& ~$\mu_R$~  & ~$\tau_R$~& ~$\overline{L}_L$~& ~$\mathcal{N}_R$~& ~$\mathcal{S}_L$~& ~$\mathcal{H}_{u,d}$~&~$\zeta$ \\ \hline 
%%%
%$SU(3)_C$ & $\bm{3}$  & $\bm{3}$ & $\bm{3}$ & $\bm{1}$ & $\bm{1}$ & $\bm{1}$ & $\bm{1}$ & $\bm{1}$ & $\bm{1}$   \\\hline 
$SU(2)_L$ & $1$  & $1$  & $1$  & $2$  & $1$  & $1$  & $2$   & $1$     \\\hline 
$U(1)_Y$   & $1$ & $1$ & $1$ & $-\frac12$  & $0$ & $0$  & $\frac12, -\frac12$  & $0$   \\\hline

%$U(1)_X$   & $1$ & $1$ & $1$   &$\red{-1}$  & $1$  & $2$  & $0$  &$1$  \\\hline
$A_5^\prime$ & $1$ & $1$ & $1$ & $3$ & $3^\prime$ & $3^\prime$ & $1$ & $1$  \\ \hline
%$k_I$ & $0$ & $0$ & $0$ & $0$ & $-2$ & $0$ & $0$ & $0$ & $-2$\\

$U(1)_{B-L}$   & $-1$ & $-1$ & $-1$   &$1$  & $-1$   &$0$ &$0$ &$1$ \\\hline

$k_I$ & $2$ & $4$ & $6$ & $0$ & $6$ & $0$ & $0$  & $0$\\ 
\hline
\end{tabular}
\caption{Particle content of the model and their charges under $SU(2)_L\times U(1)_Y\times A_5^\prime\times U_{B-L} $ group and their modular weights  $k_I$.}
\label{tab:fields-inverse}
\end{table}
\end{center}
The  superpotential of the model is given by
\begin{eqnarray}
\mathcal{W} &=&  A_{\mathcal{M}_l}\left[(\overline{L}_L l_R))_{\textbf{3}} Y^{k_I}_{\textbf{3}}\right]\mathcal{H}_d  + \mu \mathcal{H}_u \mathcal{H}_d  +  G_D \Big[(\overline{L}_L \mathcal{N}_R))_{\textbf{4}}\sum\limits^{2}_{i=1} Y_{\textbf{4},i}^{(6)}\Big] \mathcal{H}_u  \\ \nonumber
&+& B_{\mathcal{M}_{RS}} \Big [(\overline{\mathcal{S}}_L \mathcal{N}_R)_{\bf 5} \sum\limits^{2}_{i=1}Y_{\textbf{5},i}^{(6)}\Big ] \zeta  + \mu_0 \overline{S^C_L} S_L\;,
\end{eqnarray}
where, $A_{\mathcal{M}_l}$,  $G_D$ and $B_{\mathcal{M}_{RS}} $ are $3 \times 3$ diagonal matrices given as   $A_{\mathcal{M}_l}={\rm diag}\left( \alpha_{\mathcal{M}_l}, \beta_{\mathcal{M}_l}, \gamma_{\mathcal{M}_l}\right)$,   $G_D = {\rm diag} \left(g_{D_1}, g_{D_2}, g_{D_3}\right)$,  and $B_{\mathcal{M}_{RS}} = {\rm diag} \left (\alpha_{RS_1}, \alpha_{RS_2}, \alpha_{RS_3} \right)$. The modular weight $k_I$ in the first term takes the values $k_I = (2, 4, 6)$ for $l=(e,\mu,\tau)$.
%%%%%%%%%%%%%%%%%%%%%%%%%%%%%%%%%%%%%%
%\vspace{1cm}
\subsection{ Dirac mass term for charged leptons}
To establish charged leptons mass matrix, the left-handed doublet superfields i.e., $\overline{L}_{L}$, transform as triplets under the $A^\prime_5$ symmetry with $B-L$ charge $-1$. 
%Furthermore,  they are assigned as -1 under  $U(1)_{B-L}$. 
The Higgsinos $\mathcal{H}_{u,d}$ are given charges 0, 1 under the $U_{B-L}$ and $A^\prime_5$ symmetries respectively   with zero modular weight. The VEVs of these  Higgsinos  $\mathcal{H}_u$ and $\mathcal{H}_d$ are given  as $v_u/\sqrt2$ and $v_d/\sqrt2$ respectively. Moreover, Higgsinos VEVs are associated to SM Higgs VEV as $v_H = \frac12 \sqrt{v^2_u + v^2_d}$ and the ratio of their VEVs is expressed as $\tan\beta= ({v_u}/{v_d})=5$. Hence, the relevant superpotential term for charged leptons is given as
\begin{align}
 \mathcal{W}_{\mathcal{M}_l}  
%&y_{\ell_{}} \overline{L_L} \mathcal{H} \ell_R  \nonumber \\
&= \alpha_{\mathcal{M}_l}  \left[(\overline{L_L} e_R)_{\bf 3} Y_{\bf{3}}^{(2)} \right] \mathcal{H}_d +\beta_{\mathcal{M}_l}  \left[(\overline{L_L} \mu_R)_{\bf{3}} Y_{\bf{3}}^{(4)} \right] \mathcal{H}_d + \gamma_{\mathcal{M}_l}  \Big[(\overline{L_L} \tau_R)_{\bf{3}} \Big\lbrace \sum\limits_{i=1}^2Y_{{\bf{3}},i}^{(6)} \Big\rbrace \Big] \mathcal{H}_d\;.
                   % + {\rm H.c.}, 
                    \label{Eq:yuk-MD} 
 \end{align}
After the spontaneous symmetry breaking, it is evident that the charged lepton mass matrix isn't diagonal and is expressed as
\begin{align}
\mathcal{M}_{l}&=\frac{v_d}{\sqrt 6}
\left[\begin{matrix}
\left(Y^{(2)}_{\bf 3}\right)^{}_1 && \left(Y^{(4)}_{\bf 3}\right)^{}_1 && \left(\sum\limits_{i=1}^2Y^{(6)}_{{\bf 3},i}\right)_1  \\
\left(Y^{(2)}_{\bf 3}\right)^{}_3 &&\left(Y^{(4)}_{\bf 3}\right)^{}_3 &&\left(\sum\limits_{i=1}^2Y^{(6)}_{{\bf 3},i}\right)_3  \\
\left(Y^{(2)}_{\bf 3}\right)^{}_2 &&\left(Y^{(4)}_{\bf 3}\right)^{}_2 && \left(\sum\limits_{i=1}^2Y^{(6)}_{{\bf 3},i}\right)_2  \\
\end{matrix}\right]_{LR}      \cdot
\left[\begin{array}{ccc}
 \alpha_{\mathcal{M}_l}  & 0 & 0 \\ 
0 & \beta_{\mathcal{M}_l} & 0 \\ 
0 & 0 & \gamma_{\mathcal{M}_l} \\ 
\end{array}\right]\;.            
\label{Eq:Mell} 
\end{align}\\
The charged lepton mass matrix ${\mathcal M}_l$ can be diagonalised by  the unitary matrix $U_l$, giving rise to the physical masses $e,~\mu$ and $\tau$ as  
\begin{equation} 
U^\dagger_l \mathcal{M}_l \mathcal{M}^\dagger_l U_l = {\rm diag}(m^2_e, m^2_\mu, m^2_\tau)\;. 
\end{equation}
In addition, it also satisfies the following identities, which will be used for numerical analysis in section \ref{sec:results}:
\begin{eqnarray}
	{\rm Tr} \left( \mathcal{M}^{}_l \mathcal{M}^{\dag}_l \right) &=& m^{2}_{e} + m^{2}_{\mu} + m^{2}_{\tau} \; ,  \nonumber\\
	{\rm Det}\left( \mathcal{M}^{}_l \mathcal{M}^{\dag}_l \right) &=& m^{2}_{e} m^{2}_{\mu} m^{2}_{\tau} \; , \nonumber\\
	\dfrac{1}{2}\left[{\rm Tr} \left(\mathcal{M}^{}_l \mathcal{M}^{\dag}_l\right)\right]^2_{} - \dfrac{1}{2}{\rm Tr}\left[ (\mathcal{M}^{}_l \mathcal{M}^{\dag}_l)^2_{}\right] &= & m^{2}_{e}m^{2}_{\mu}+m^{2}_{\mu}m^{2}_{\tau}+m^{2}_{\tau}m^{2}_{e} \; . \label{eq:tr}
	%    (3.4-3.5)
	\end{eqnarray}

%\noindent \\
%{\bf Dirac mass term for the small neutrinos:}
\subsection{Dirac mass term for neutrinos}
%\vspace{3mm}
%As discussed earlier about the SM lepton doublet superfields transformation, 
The right-handed neutrino superfields ${\cal N}_{R_i}$ are $\bm{3^\prime}$ under $A^\prime_5$ modular group with a ${B-L}$ charge of $-1$ and modular weight $6$.  Therefore, the invariant  superpotential, 
describing the Dirac  mass term for the  neutrinos can be written as,
\begin{align}
 \mathcal{W}_{D}  
%&y_{\ell_{}} \overline{L}_L H \ell_R  \nonumber \\
&=   G_D  \left[(\overline{L}_L~\mathcal{N}_R)_\textbf{4} \sum\limits^{2}_{i=1} Y_{\textbf{4},i}^{(6)} \right] \mathcal{H}_u\;.
       %             + {\rm H.c.}, 
                    \label{Eq:yuk-MD} 
\end{align}
Here, the subscript for the operator $\overline{L}_L \mathcal{N}_R$ indicates $A^\prime_5$ representation constructed by the Kronecker product rule (see Appendix \ref{prod_rules}) which further leads in obtaining a invariant superpotential. The resulting Dirac neutrino mass matrix is found to be
\begin{align}
\mathcal{M}_D&=\frac{v_u}{2\sqrt6} 
\left[\begin{array}{ccc}
0 & -\sqrt2 \left(\sum\limits^{2}_{i=1} Y_{\textbf{4},i}^{(6)}\right)_{{3}} & -\sqrt2 \left(\sum\limits^{2}_{i=1} Y_{\textbf{4},i}^{(6)}\right)_{{2}} \vspace{2mm}\\ 
\sqrt2 \left(\sum\limits^{2}_{i=1} Y_{\textbf{4},i}^{(6)}\right)_{{4}} &  \left(\sum\limits^{2}_{i=1} Y_{\textbf{4},i}^{(6)}\right)_{{2}} &-\sqrt2 \left(\sum\limits^{2}_{i=1} Y_{\textbf{4},i}^{(6)}\right)_{{1}}\vspace{2mm} \\ 
\sqrt2 \left(\sum\limits^{2}_{i=1} Y_{\textbf{4},i}^{(6)}\right)_{{1}} & \left(\sum\limits^{2}_{i=1} Y_{\textbf{4},i}^{(6)}\right)_{{4}} & -\left(\sum\limits^{2}_{i=1} Y_{\textbf{4},i}^{(6)}\right)_{{3}} \\ 
\end{array}\right]_{LR}.
\begin{bmatrix}
g_{D_1} & 0 & 0 \\
0 & g_{D_2} & 0 \\
0 & 0 & g_{D_3}
\end{bmatrix}\;,         
\label{Eq:Mell} 
\end{align}
where $(g_{D_1}, g_{D_2}, g_{D_3})$ are the free parameters of the diagonal matrix $G_D$.
%===================
%{\bf Mixing between the heavy fermions $N_R$ and $S_L$:}
%\vspace{2cm}
\subsection{ Mixing between the heavy fermions $\mathcal{N}_R$ and $\mathcal{S}_L$}
%\vspace{3mm}
The mixing between heavy fermion superfields ${\cal N}_R$ and ${\cal S}_L$   can be expressed as follows,
\begin{eqnarray}
 \mathcal{W}_{\mathcal{M}_{RS}}  
%&y_{\ell_{}} \overline{L_L} H \ell_R  \nonumber \\
                  &=& B_{\mathcal{M}_{RS}} \Big [(\overline{\mathcal{S}}_L \mathcal{N}_R)_{\bf 5} \sum\limits^{2}_{i=1}Y_{\bf 5,i}^{(6)}\Big ] \zeta
%                  + \beta_{NS} \bm{Y} (\overline{S_L} N_R)_{\rm Anti-symm} ]\zeta   + {\rm H.c.} 
%                   &=&\alpha_{NS}[ y_1(2  \bar S_{L_1} N_{R_1} - \bar S_{L_2} N_{R_3} - \bar S_{L_3} N_{R_2})+y_2(2  \bar S_{L_2} N_{R_2} - \bar S_{L_1} N_{R_3} - \bar S_{L_3} N_{R_1}) \nn \\
%&&+ y_3(2  \bar S_{L_3} N_{R_3} - \bar S_{L_1} N_{R_2} - \bar S_{L_2} N_{R_1})] \zeta \nn\\
%&&+\beta_{NS}[ y_1(  \bar S_{L_2} N_{R_3} - \bar S_{L_3} N_{R_2})+y_2(  \bar S_{L_3} N_{R_1} - \bar S_{L_1} N_{R_3})+
%y_3(  \bar S_{L_1} N_{R_2} - \bar S_{L_2} N_{R_1})] \zeta  \nn \\ 
%&& + {\rm H.c.},             
      \label{Eq:yuk-M} 
\end{eqnarray}
where, the choice of Yukawa coupling depends on the sum of the modular weight of the superfields and the Kronecker product rule as given in  Appendix \ref{prod_rules}.  Using $\langle \zeta \rangle = v_\zeta/\sqrt{2}$,  the resulting mass matrix is found to be
\begin{align}
\mathcal{M}_{RS}&=\frac{v_\zeta}{2\sqrt15}
\left[\begin{array}{ccc}
2 \left(\sum\limits_{i=1}^2 Y_\textbf{5,i}^{(6)}\right)_1 & -\sqrt3\left(\sum\limits_{i=1}^2 Y_\textbf{5,i}^{(6)}\right)_4 & -\sqrt3\left(\sum\limits_{i=1}^2 Y_\textbf{5,i}^{(6)}\right)_3 \\ 
 -\sqrt3\left(\sum\limits_{i=1}^2 Y_\textbf{5,i}^{(6)}\right)_4 &-\sqrt6  \left(\sum\limits_{i=1}^2 Y_\textbf{5,i}^{(6)}\right)_2 &-\left(\sum\limits_{i=1}^2 Y_\textbf{5,i}^{(6)}\right)_1 \\ 
-\sqrt3 \left(\sum\limits_{i=1}^2 Y_\textbf{5,i}^{(6)}\right)_3 &-\left(\sum\limits_{i=1}^2 Y_\textbf{5,i}^{(6)}\right)_1 &-\sqrt6  \left(\sum\limits_{i=1}^2 Y_\textbf{5,i}^{(6)}\right)_5 \\ 
\end{array}\right]_{LR}. \begin{bmatrix}
\alpha_{RS_1} & 0 & 0 \\
0 & \alpha_{RS_2} & 0 \\
0 & 0 & \alpha_{RS_3}
\end{bmatrix}\;, \label{yuk:MRS}
\end{align}
where  $(\alpha_{RS_1}, \alpha_{RS_2}, \alpha_{RS_3})$ are the free paramaters of the  diagonal matrix $B_{\mathcal{M}_{RS}}$.

\subsection{Majorana mass term for $\mathcal{S}_L$}
Under $A^\prime_5$ singlet heavy fermions ${\cal S}_L$ transform as triplet $\bm{3^\prime}$ having zero modular weight. Hence, its Majorana mass term can be written as,
\begin{align}
 \mathcal{W}_{\mu}  
%&y_{\ell_{}} \overline{L_L} H \ell_R  \nonumber \\
                   &= \mu_0 \mathcal{S}_L \mathcal{S}_L  , \label{Eq:yuk-M}
\end{align}
leading to the mass matrix ($\mathcal{M}_\mu$) of the form
\begin{align}
\mathcal{M}_\mu = \mu_0 \begin{bmatrix}
1 & 0 & 0 \\
0 & 0 & 1 \\
0 & 1 & 0
\end{bmatrix}.
\end{align}

\subsection{Inverse Seesaw mechanism for light neutrino Masses}
In the present model constructed using $A^\prime_5$ modular symmetry, the complete $9 \times 9$ neutral fermion mass 
matrix  in the flavor basis of $\left(\nu_L, {\cal N}_R, {\cal S}^c_L \right)^T$ is given as
\begin{eqnarray}
\mathbb{M} = \left(\begin{array}{c|ccc}   & \nu_L & \mathcal{N}_R  & \mathcal{S}^c_L   \\ \hline
\nu_L  & 0       & \mathcal{M}_D       & 0 \\
\mathcal{N}_R    & \mathcal{M}^T_D         & 0       & \mathcal{M}_{RS} \\
\mathcal{S}^c_L & 0     & \mathcal{M}_{RS}^T    & \mathcal{M}_\mu
\end{array}
\right).
\label{eq:numatrix-complete}
\end{eqnarray}
 In the limit $ \mathcal{M}_\mu \ll M_{D} < M_{RS}$,   the above mass matrix (\ref{eq:numatrix-complete})  provides the  inverse seesaw mass formula for the light neutrinos as
\begin{eqnarray}
m_\nu 
&=& \mathcal{M}_D~ \mathcal{M}_{RS}^{-1} ~\mathcal{M}_\mu~ (\mathcal{M}_{RS}^{-1})^T~ (\mathcal{M}_D)^T .\label{mass}
%\left(\frac{M_D}{M_{NS}}\right) \mu \left(\frac{M_D}{M_{NS}}\right)^T\,. 
\end{eqnarray}
Thus, diagonalization of the light neutrino mass matrix (\ref{mass}) yields the masses of the active neutrinos.  Apart from determining the  small neutrino masses,  other parameters, which are of great use, are the Jarlskog invariant ($J_{CP}$) and the effective neutrino mass $\langle m_{ee} \rangle$ describing the neutrinoless double beta decay. These parameters related  to the mixing angles and phases of PMNS matrix through
\begin{eqnarray}
&& J_{CP} = \text{Im} [U_{e1} U_{\mu 2} U_{e 2}^* U_{\mu 1}^*] = s_{23} c_{23} s_{12} c_{12} s_{13} c^2_{13} \sin \delta_{CP}\;,\\
&& \langle m_{ee}\rangle=|m_{\nu_1} \cos^2\theta_{12} \cos^2\theta_{13}+ m_{\nu_2} \sin^2\theta_{12} \cos^2\theta_{13}e^{i\alpha_{21}}+  m_{\nu_3} \sin^2\theta_{13}e^{i(\alpha_{31}-2\delta_{CP})}|.
\end{eqnarray}
The effective Majorana mass parameter $\langle m_{ee}\rangle$ is expected to have improved sensitivity measured by KamLAND-Zen experiment in coming future~\cite{KamLAND-Zen:2016pfg}. 
\section{NUMERICAL ANALYSIS}
\label{sec:results}
Numerical analysis is performed by considering experimental data at 3$\sigma$ interval~\cite{deSalas:2020pgw} as follows:
\begin{align}
&{\rm NO}: \Delta m^2_{\rm atm}=[2.47, 2.63]\times 10^{-3}\ {\rm eV}^2,\
\Delta m^2_{\rm sol}=[6.94, 8.14]\times 10^{-5}\ {\rm eV}^2,\nn\\
&\sin^2\theta_{13}=[0.0200, 0.02405],\ 
\sin^2\theta_{23}=[0.434, 0.610],\ 
\sin^2\theta_{12}=[0.271, 0.369]\;. \label{eq:mix}
%%% 
\end{align} 
Here, numerical diagonalization of the light neutrino mass matrix as given in eqn.(\ref{mass}) is done through $U_\nu^\dagger {\cal M}U_\nu= {\rm diag}(m_1^2, m_2^2, m_3^2)$, where  ${\cal M}=m_\nu m_\nu^\dagger$ and $U_\nu$ is an unitary matrix. Thus, the lepton mixing matrix is given as  $U=U_l^\dagger U_\nu$, from which the neutrino mixing angles can be extracted  using the standard relations:
\begin{eqnarray}
\sin^2 \theta_{13}= |U_{13}|^2\;,~~~~\sin^2 \theta_{12}= \frac{|U_{12}|^2}{1-|U_{13}|^2}\;,~~~~~\sin^2 \theta_{23}= \frac{|U_{23}|^2}{1-|U_{13}|^2}\;.
\label{mixangles}
\end{eqnarray}
In order to demonstrate the current neutrino oscillation data,  the  values of model parameters are chosen to be in the following ranges:
\begin{align}
&{\rm Re}[\tau] \in [0, 0.5],~~{\rm Im}[\tau]\in [0.5, 2],~~\{ g_{D_1},g_{D_2},g_{D_3} \} \in ~[10^{-4},10^{-2}], \nn \\
&  \{\alpha_{RS_1}, \alpha_{RS_2}, \alpha_{RS_3}\}  \in [0.1, 1],\quad v_\zeta \in \nn [1,100] \ {\rm TeV}\;  .
\end{align}
For diagonalizing the charged lepton mass matrix ${\mathcal M}_l$ (\ref{Eq:Mell}), we use  the values of the free parameters  as: $\alpha_{\mathcal{M}_l} \approx \mathcal{O}(10^{-6})$, $\beta_{\mathcal{M}_l} \approx \mathcal{O}(10^{-2})$ and $\gamma_{\mathcal{M}_l} \approx  \mathcal{O}(10^{-4})$, and scanning over the the allowed ranges of real and imaginary parts of the modulus $\tau$, i.e., 0\ $\lesssim\ $Re$[\tau]\lesssim$\ 0.5 and  0.5\ $\lesssim\ $Im$[\tau]\lesssim$\ 2 , we numerically obtain the diagonalizing matrix $U_l$, that gives  the charged-lepton masses  as $m_e = 0.511~ \rm{MeV}$, $m_\mu = 105.66~ \rm{MeV}$, $m_\tau = 1776.86~ \rm{MeV}$. 

%%%%%%%%%%%%%%%%%%%%%%%%%%%%%%%%%%% 
\begin{figure}[h!]
\begin{center}
\includegraphics[height=50mm,width=75mm]{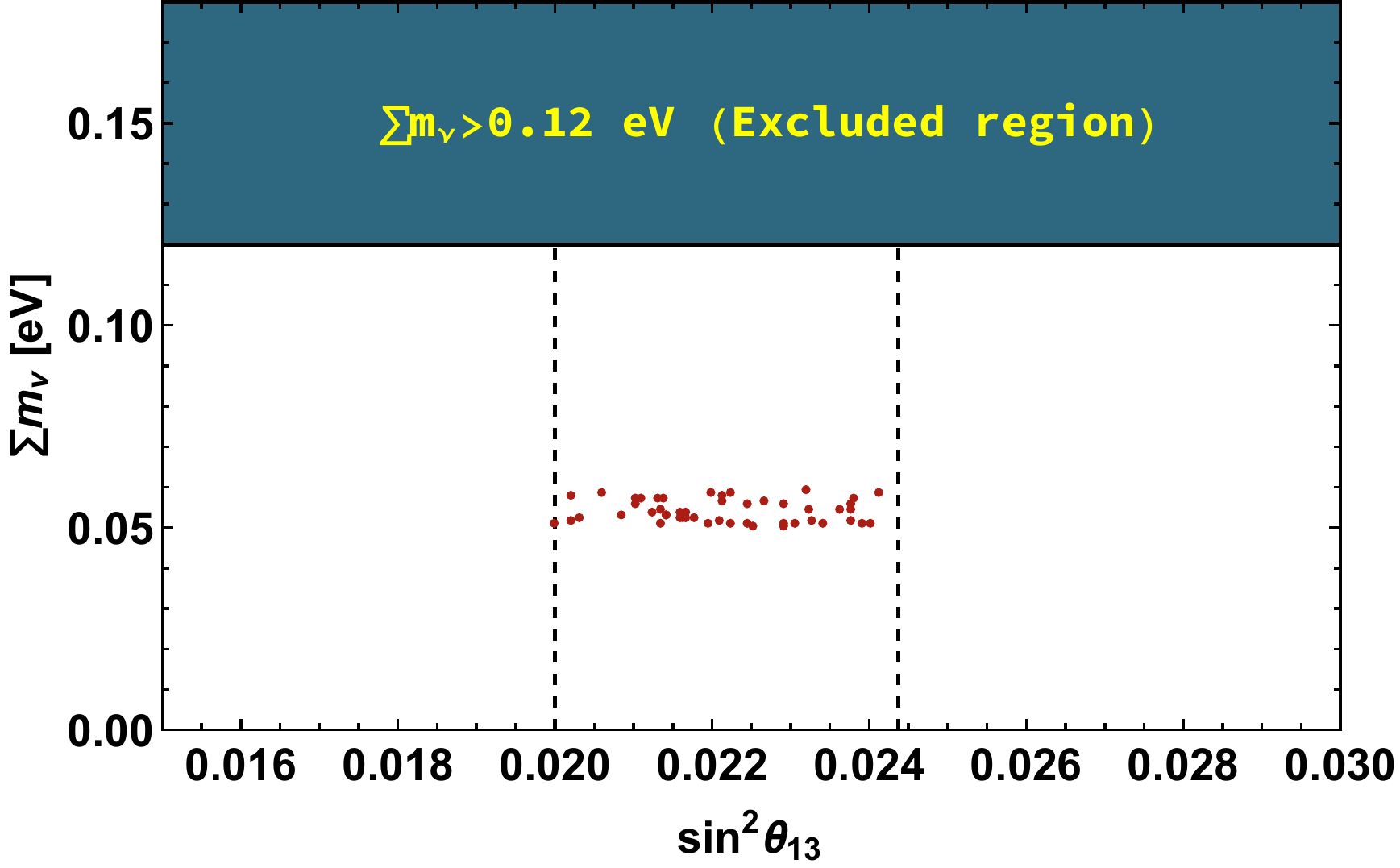}
\hspace*{0.2 true cm}
\includegraphics[height=50mm,width=75mm]{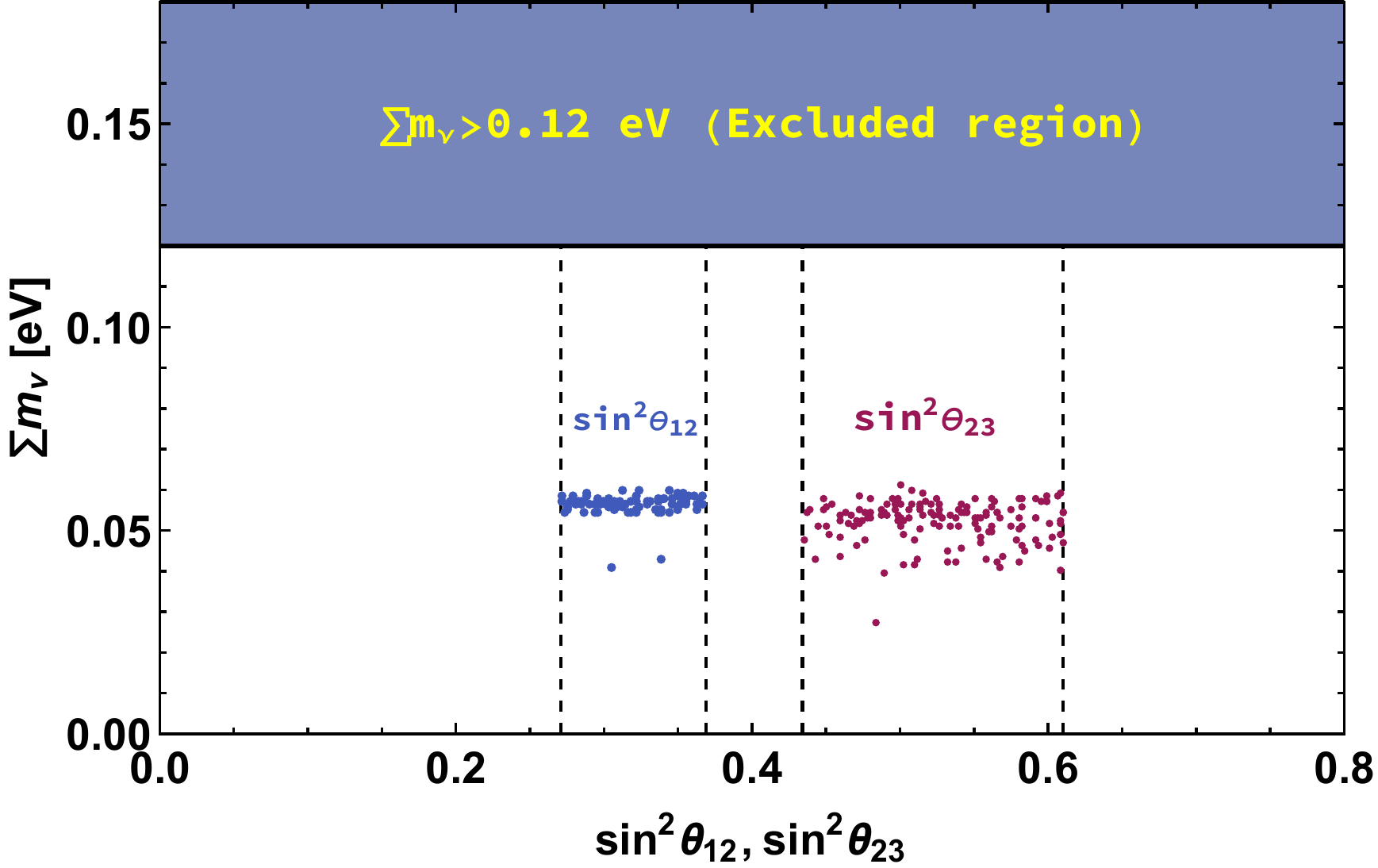}
\caption{Left (right) panel signify the correlation of the mixing angles i.e. $\sin^2 \theta_{13}$ ($\sin^2 \theta_{12}$, $\sin^2 \theta_{23}$) respectively with the sum of neutrino masses $\sum m_i$ (eV). }
\label{nmass_mix}
\end{center}
\end{figure}
\begin{figure}[h!]
\begin{center}
\includegraphics[height=50mm,width=75mm]{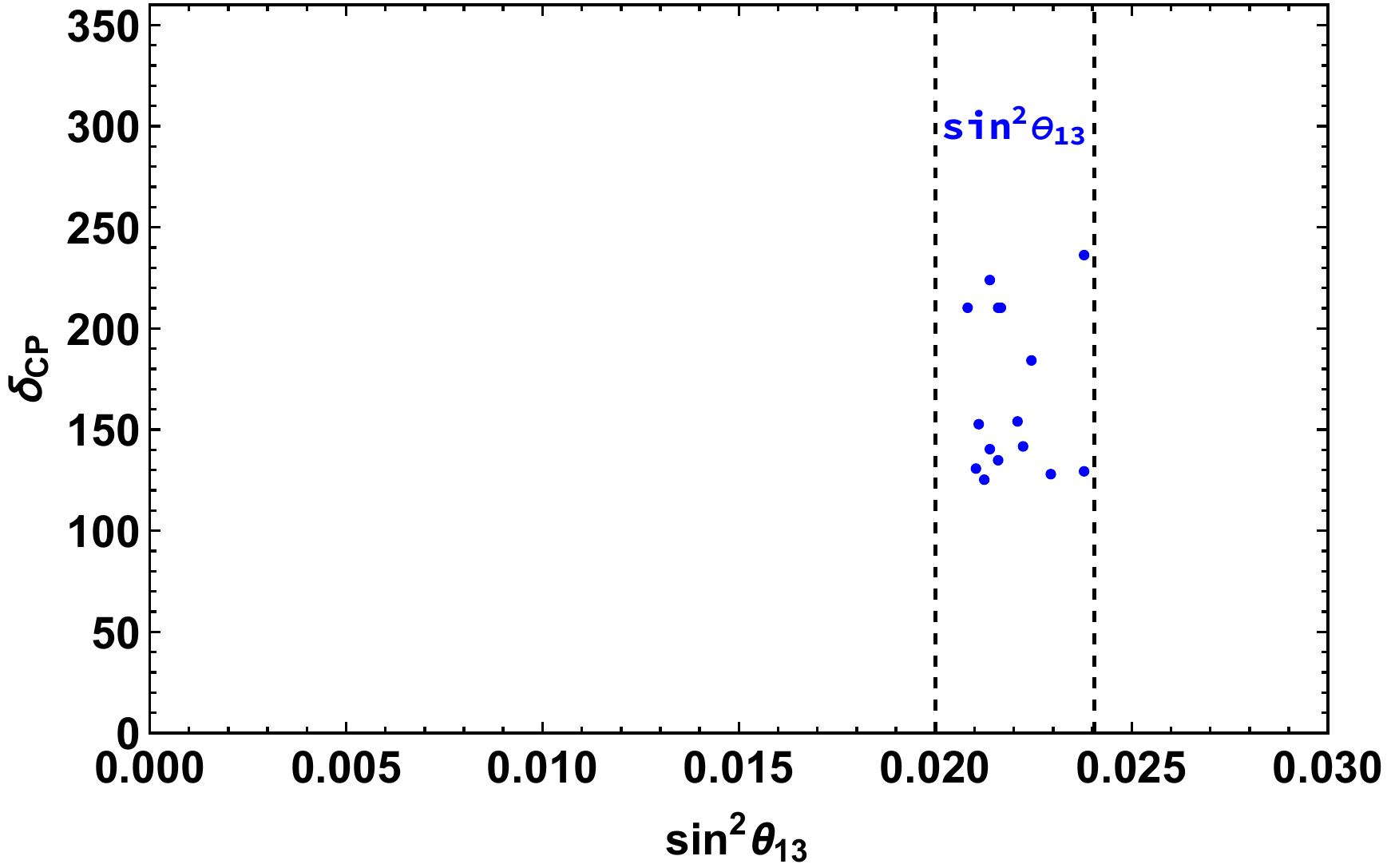}
\hspace*{0.2 true cm}
\includegraphics[height=50mm,width=75mm]{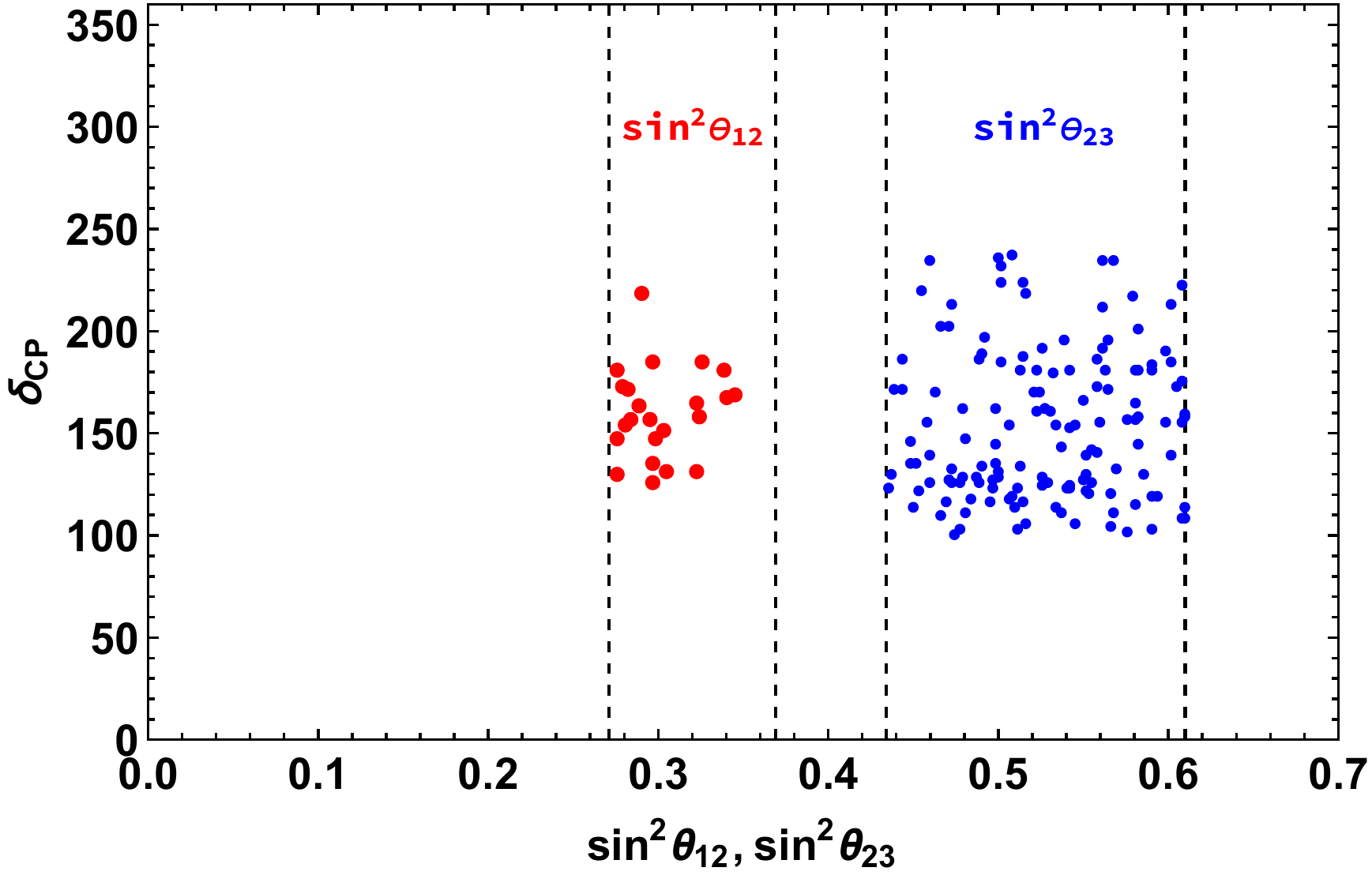}
\caption{Left (right) panel displays the correlation between $\delta_{CP}$ w.r.t $sin^2\theta_{13}$ ($sin^2\theta_{12}$ and $sin^2\theta_{23}$).}
\label{deltacp_mixangle}
\end{center}
\end{figure}
\begin{figure}[h!]
\begin{center}
\includegraphics[height=50mm,width=75mm]{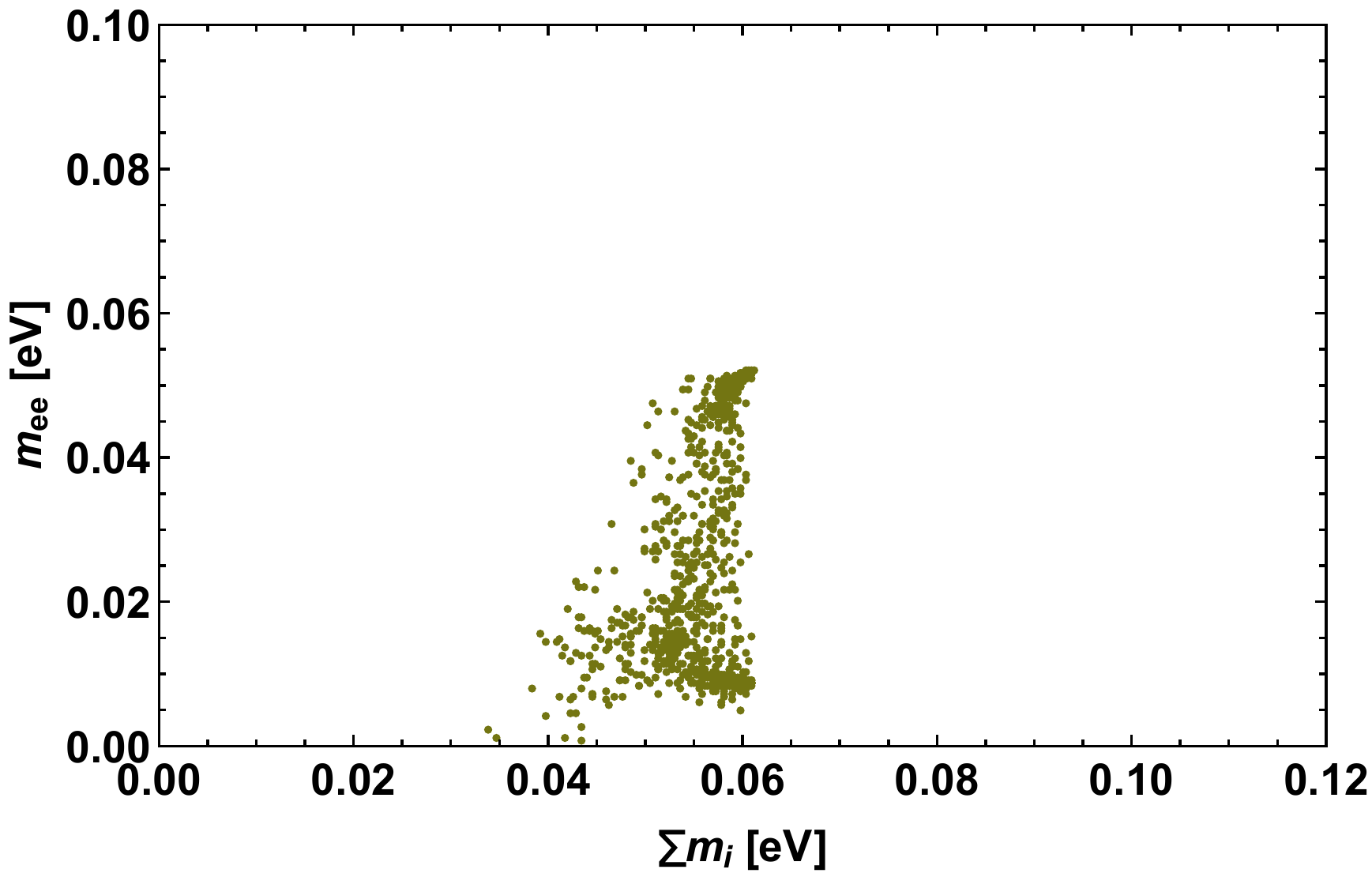}
\hspace*{0.2 true cm}
\includegraphics[height=48mm,width=72mm]{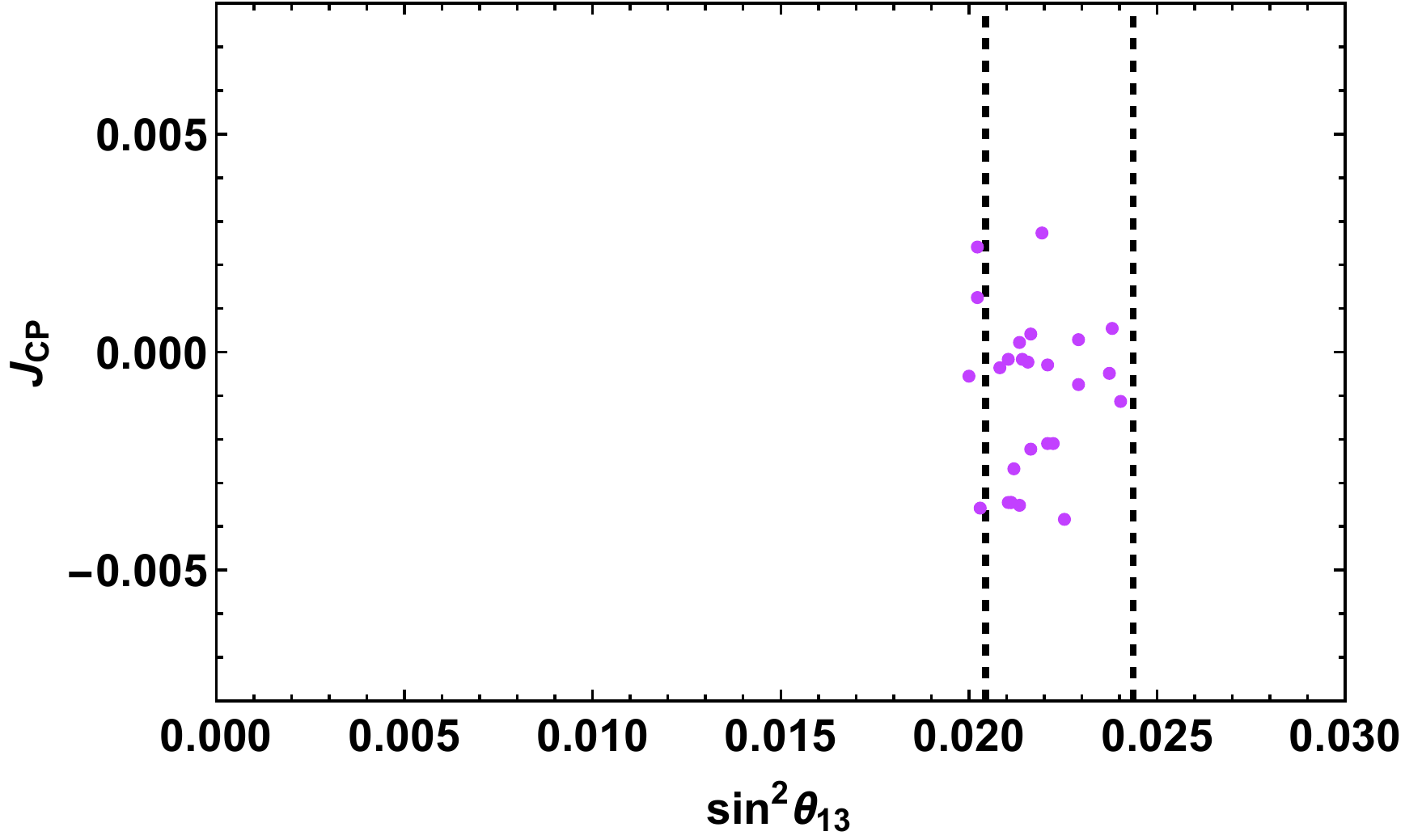}
\caption{Left panel shows   the correlation of effective neutrino mass of neutrinoless double beta decay with the sum of active neutrino masses and right panel represents a correlation between $\rm{J_{CP}}$ with respect to the reactor mixing angle.}
\label{meeandjcpss13}
\end{center}
\end{figure}
\begin{figure}[h!]
\begin{center}
\includegraphics[height=50mm,width=75mm]{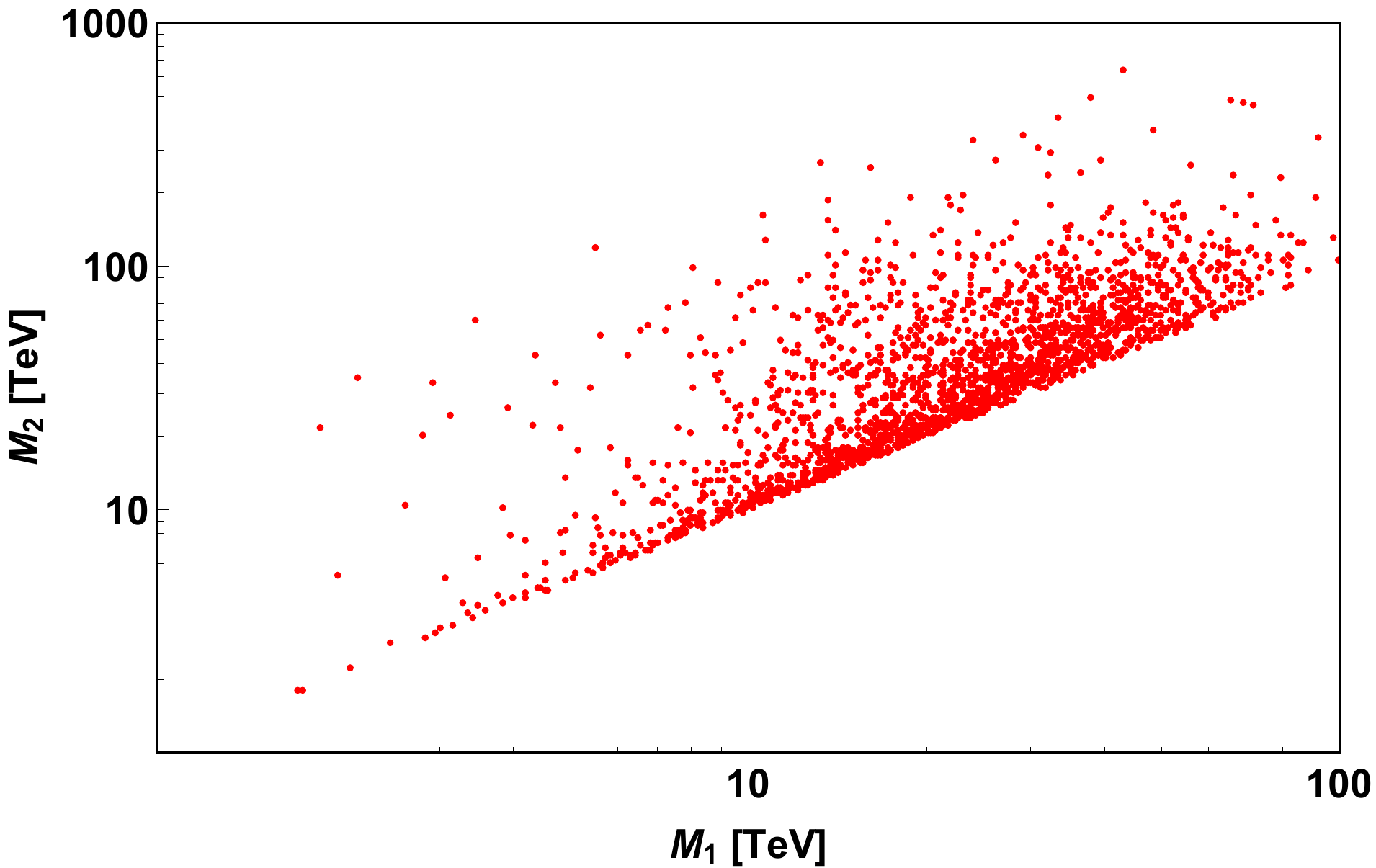}
\hspace*{0.2 true cm}
\includegraphics[height=48mm,width=72mm]{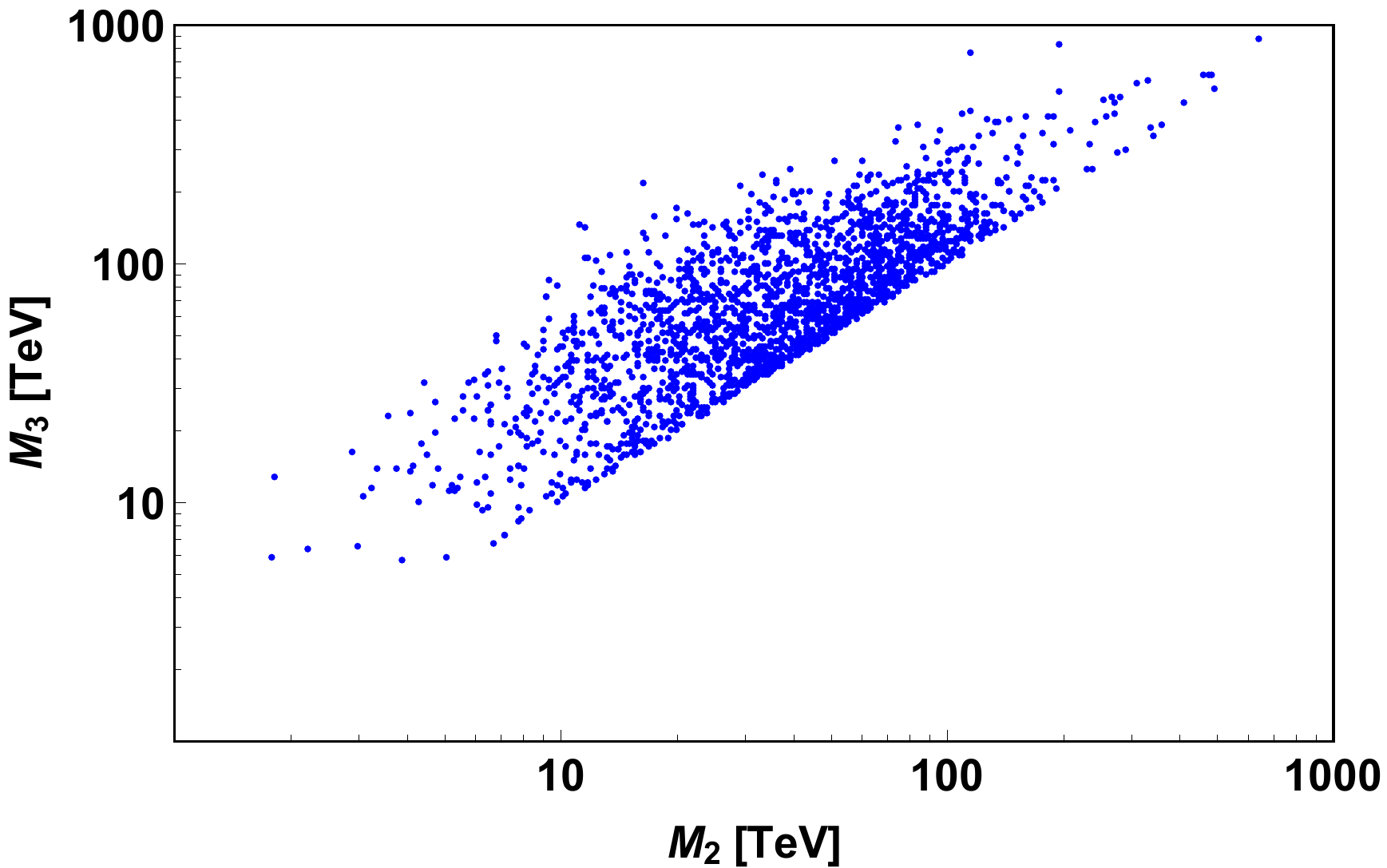}
\caption{Left panel shows the correlation of heavy fermion masses $M_1$ versus $M_2$ and right panel represents a correlation of heavy fermion masses $M_2$ versus $M_3$ in TeV scale.}
\label{m1m2m3}
\end{center}
\end{figure}

%%%%%%%%%%%%%%%%%%%%%%%%%%%%%%%%%%%%%%%%
In order to make appropriate predictions of the neutrino mixing angles and other parameters within their $3\sigma$ ranges, the input parameters are generated in a random fashion. The allowed ranges of solar and atmospheric mass squared differences at $3\sigma$ level used as constraints to calculate other neutrino oscillation parameters in their $3\sigma$ ranges \cite{deSalas:2020pgw}. Here, we have kept the range of modulus $\tau$ as:  0\ $\lesssim\ $Re$[\tau]\lesssim$\ 0.5 and  0.5\ $\lesssim\ $Im$[\tau]\lesssim$\ 2  and also the estimated range for $\mu_0 \in [10^{-5},10]~\rm keV$ for obtaining the  neutrino masses in normal ordering (NO).  With these values, the neutrino mixing angles are then extracted  using eqn. (\ref{mixangles}). The variation of the mixing angles $\sin^2{\theta_{13}}$ (left panel)  and $\sin^2{\theta_{12}}$, $\sin^2{\theta_{23}}$ (right panel) with respect to sum of the active neutrino masses are shown in Fig. \ref{nmass_mix}. Further, the variation of $\delta_{CP}$ with respect to mixing angles $\sin^2{\theta_{13}}$ (left panel)  and $\sin^2{\theta_{12}}$, $\sin^2{\theta_{23}}$ (right panel) is shown in Fig. \ref{deltacp_mixangle}, where the vertical dashed lines represent the  in $3\sigma$ ranges of the mixing angles.  The left panel of Fig. \ref{meeandjcpss13}, signifies the correlation between the  observed sum of active neutrino masses ($\sum m_i$) and the effective neutrinoless double beta decay mass parameter ($m_{ee}$) whose maximum value is found to around 0.06 eV. In the right panel of Fig. \ref{meeandjcpss13}, we  show the correlation of Jarsklog CP invariant allowed by the neutrino data, with the reactor mixing angle, which is found to be of the order of ${\cal O}(10^{-3})$. In Fig.\ref{m1m2m3} we represent the correlations between the heavy fermion masses, where, left panel is the plot expressing $M_1$ with $M_2$ and right panel is of $M_2$ versus $M_3$ in TeV scale.

%%%%%%%%%%%%%%%%%%%%%%%%%%%%%%%%%%%%%%%
\vspace{-1cm}
%%%%%%%%%%%%%%%%%%%%%%%%%%%%%%%%%%%%%%%%
\vspace{-5mm}
%%%%%%%%%%%%%%%%%%%%%%%%%%%%%%%%%%%%%%%%
%%%%%%%%%%%%%%%%%%%%%%%%%%%%%%%%%%%%%%%%
%\newpage
\section{Comments on non-unitarity}
\label{sec:non-unitarity}
In the section, we briefly comment on non-unitarity of neutrino mixing matrix $U'_{\rm PMNS}$. The form for the deviation from unitarity is expressed as following  \cite{Forero:2011pc}
\begin{align}
U'_{\rm PMNS}\equiv \left(1-\frac12 \mathcal{F}\mathcal{F}^\dag\right) U_{\rm PMNS}\;.
\end{align}
Here  $\rm{U_{PMNS}}$ is the PMNS mixing matrix, used in diagonalising the mass matrix of the three light neutrinos and $\mathcal{F}$ represents the mixing of active neutrinos with the heavy fermions and its form is given by $\mathcal{F}\equiv  (\mathcal{M}^{T}_{RS})^{-1} \mathcal{M}_D \approx ({g_D v_u}/{\alpha_{RS} v_\zeta})$, which is hermitian in nature.
The global constraints on the non-unitarity parameters \cite{Antusch:2014woa,Blennow:2016jkn,Fernandez-Martinez:2016lgt}, come from several experimental results such as the $W$ boson mass $M_W$, the  Weinberg angle $\theta_W$, several ratios of  fermionic $Z$ boson as well as its  invisible decay, electroweak universality,  CKM unitarity bounds, and lepton flavor violations. In the context of the present model, we consider the following approximated mass values for the Dirac,  Majorana mass for ${\cal S}_L$ and the pseudo-Dirac mass for the heavy fermions to correctly generate the observed mass square differences of the desired order as:
\begin{eqnarray}
\left(\frac{m_\nu}{0.1 ~{\rm eV}} \right) \approx \left(\frac{\mathcal{M}_D}{10^{2}~~ {\rm GeV}}\right)^{2} \left(\frac{{\cal M}_\mu}{ {\rm keV}}\right) \left( \frac{\mathcal{M}_{RS}}{10^{4}~~ {\rm GeV}}\right)^{-2}. 
\label{order}
\end{eqnarray}
Using the benchmark to obtain the correct order of neutrino mass as shown in eqn.(\ref{order}). Therefore, the approximated non-unitary mixing for the present model is given below: 
\begin{align}
|\mathcal{F}\mathcal{F}^\dag|\le  
\left[\begin{array}{ccc} 
1.1\times 10^{-5} & 8.3\times 10^{-7}  & 3.8\times 10^{-6}  \\
8.3\times 10^{-7}  & 9.5\times 10^{-8}  & 5.02\times 10^{-7}  \\
3.8\times 10^{-6}  & 5.02\times 10^{-7}  & 3.05\times 10^{-7} \\
 \end{array}\right]\;.
 \label{non-uni}
\end{align}

%\blue{Since, the non-unitarity can be controlled by the $v_\zeta$ as it is of large mass scale. Hence, the results are expected as represented above.}
%%%%%%%%%%%%%%%%%%%%%%
\section{comments on LFV}
\label{sec:LFV}
Lepton flavour violation is one of the most fascinating probes for new physics beyond the SM, therefore, here we investigate decay mode $\mu \to e \gamma$. Several experiments are looking for this decay mode  with great effort for an improved sensitivity, and the current  limit on its branching  ratio is from MEG collaboration as Br$(\mu\rightarrow e\gamma)< 4.2\times 10^{-13}$ \cite{TheMEG:2016wtm}. There is a sizeable contribution in the present model using the $A^\prime_5$ inverse seesaw mechanism, due to the allowed light-heavy neutrino mixing. The branching ratio for the $\mu \to e\gamma$ in our model framework is given by 
\begin{align}
	{\rm Br}(\mu \to e \gamma) = 
		\left[\frac{3}{16}\right] \left[\frac{\alpha}{2\pi}\right] \sum_{i=1}^3 f\left(\frac{\mathcal{M}^2_i}{\mathcal{M}^2_W}\right) 
		\left| \mathcal{F}_{\mu i}^\ast \, \mathcal{F} _{e i}\right|^2 \,.
\end{align}
Here, $M_i$ represents the heavy fermions mass and $f(M^2_i/M^2_W)$ is a loop-function \cite{Ibarra:2011xn}. Also  $\mathcal{F} \equiv  (\mathcal{M}^{T}_{RS})^{-1} \mathcal{M}_D=({v_u}/{v_\zeta})(\tilde{\mathcal{M}}^{T}_{RS})^{-1} \tilde{\mathcal{M}}_D \approx ({v_u}/{v_\zeta})$ when $\mathcal{M}_{D}$ and $\mathcal{M}_{RS}$ are of same order. 
\begin{figure}[h!]
\begin{center}
\includegraphics[height=50mm,width=75mm]{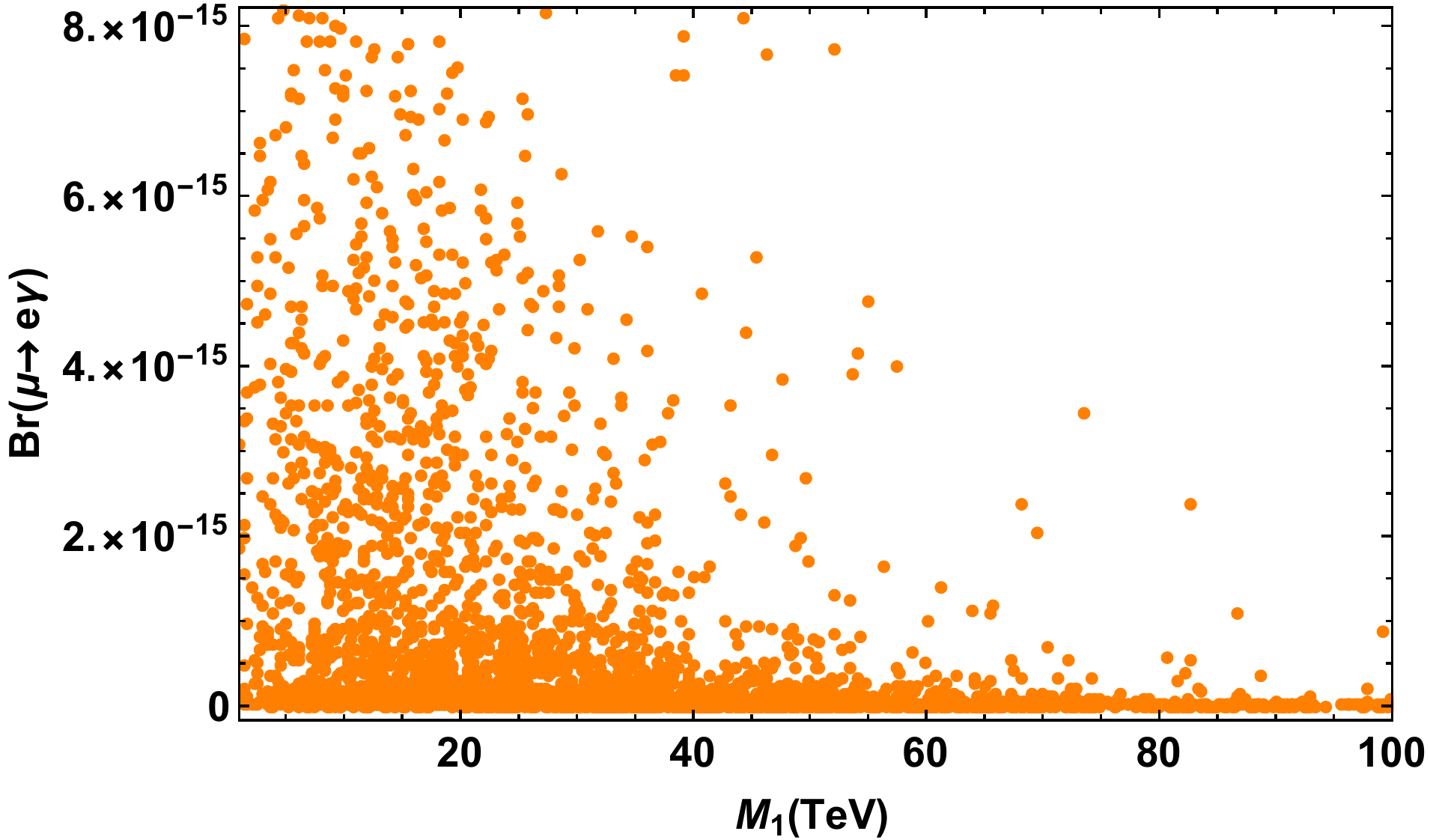}
\caption{Plot above represents the correlation between ${\rm Br}(\mu \to e \gamma)$ with respect to (lightest heavy fermion) $M_1$ in TeV scale.}
\label{LFV }
\end{center}
\end{figure}
The  branching ratio plot for the  lepton flavor violating decay $\mu \to e \gamma$ is presented against lightest heavy fermion mixing mass $M_1$ is shown in Fig.  \ref{LFV }. 
From the figure,  it is evident that the predicted branching ratio is well below the current upper limit mentioned above. 
\section{Conclusion}
\label{sec:con}
We have investigated the implications of modular  $A^\prime_5$ flavor symmetry on neutrino phenomenology The current model  three right-handed and three left handed heavy neutral fermions to incorporate the inverse seesaw framework. The singlet scalar $\zeta$ played an imperative role in spontaneous breaking of $U(1)_{B-L}$ local symmetry and gave masses to the heavy fermions. We have considered higher order Yukawa couplings that obey the rule $k^{}_{Y} = k^{}_{I^{}_1} + k^{}_{I^{}_2} + \cdots + k^{}_{I^{}_n}$, where $k^{}_{Y}$ is the weight on the Yukawa coupling and $k^{}_{I^{}_i} (i=1,2,3,4\cdots)$ are the weights on the superfields under $A^\prime_5$ symmetry. This helped us to attain a specific flavor structure for neutrino mass matrix. Proceeding further, we diagonalize the mass matrix numerically and are able to vary the model parameters in such a way that they yield results compatible to $3\sigma$ limit of oscillation data. In addition, we also investigated lepton flavour violating decay mode  $\mu\to e\gamma$ and found that its predicted  branching ratio is well below the  present experimental upper limit $4.2 \times10^{-13}$.

\acknowledgments

MKB want  to acknowledge DST for its financial help. RM  acknowledges the support from  SERB, Government of India, through grant No. EMR/2017/001448 and University of Hyderabad IoE project grant no. RC1-20-012.

% BIBLIOGRAPHY
% use BIBTEX if you want
%\bibliographystyle{JHEP}
%\bibliography{yourBIBfiles}


 %%%%%%%%%%%%%%%%%%%%%%%%%%%%%%%%%%%%%%%%%%%%%%%%%%%%%%%%%%%
\appendix
\section{The modular space of $\Gamma(5)$}
\label{mod_space}
In order to establish the modular forms which transform nontrivially under $\Gamma^{\prime}_{5}$, and are isomorphic to $A^\prime_5$, it is first required  to find out the modular space of $\Gamma(5)$. Hence, if $k$ is an integer i.e. non-negative, the modular space ${M}^{}_{k}\left[\Gamma(5)\right]$ bearing weight $k$ for $\Gamma(5)$ contains $5k+1$ linearly independent modular forms, which act like the basis vectors of the modular space. Thus, one can have
\begin{eqnarray}
{M}^{}_k\left[\Gamma(5)\right] = \bigoplus^{}_{\substack{a+b = 5k \\a, b \geq 0}} \mathbb{C}\, \frac{\eta(5 \tau)^{15 k}}{\eta(\tau)^{3 k}} \, {\mathfrak k}^a_{\frac{1}{5},\frac{0}{5}}(5\tau) \, {\mathfrak k}^b_{\frac{2}{5},\frac{0}{5}}(5\tau) \; ,
\label{eq:G5basis}
%     (2.11)
\end{eqnarray}
below given is the Dedekind eta function $\eta(\tau)$
\begin{eqnarray}
\eta(\tau)=q^{1 / 24}_{} \prod_{n=1}^{\infty}\left(1-q^{n}_{}\right) \; ,
\label{eq:Dedekindeta}
%     (2.12)
\end{eqnarray}
where, $q \equiv e^{2 {\rm i} \pi \tau}_{}$, and ${\mathfrak k}^{}_{r^{}_1,r^{}_2}(\tau)$ is the Klein form
\begin{eqnarray}
\mathfrak{k}^{}_{r^{}_{1}, r^{}_{2}}(\tau)= q_{z}^{(r^{}_{1}-1) / 2}\left(1-q^{}_{z}\right) \times \prod_{n=1}^{\infty}\left(1-q^{n}_{} q^{}_{z}\right)\left(1-q^{n}_{} q_{z}^{-1}\right)\left(1-q^{n}_{}\right)^{-2}_{} \; ,
\label{eq:Kexpansion}
%     (2.13)
\end{eqnarray}
where $(r^{}_1,r^{}_2)$ depicts a pair of rational numbers in the domain of ${\mathbb Q}^2_{}-{\mathbb Z}^2_{}$, $z \equiv \tau r^{}_{1} + r^{}_{2}$ and $q^{}_z \equiv e^{2 {\rm i} \pi z}$. Under the transformations of $S$ and $T$, the eta function and the Klein form change as follows
\begin{eqnarray}
\begin{array}{cclcl}
S &: ~\quad & \eta(\tau) \rightarrow \sqrt{-{\rm i} \tau} \eta(\tau) \; , &~\quad &\mathfrak{k}^{}_{r^{}_{1}, r^{}_{2}}(\tau) \rightarrow -\dfrac{1}{\tau}\, \mathfrak{k}^{}_{-r^{}_{2}, r^{}_{1}}(\tau) \; , \\
T &: ~\quad &\eta(\tau) \rightarrow e^{\rm{i} \pi / 12}_{} \eta(\tau) \; , & ~\quad &\mathfrak{k}^{}_{r^{}_{1}, r^{}_{2}}(\tau) \rightarrow \mathfrak{k}^{}_{r^{}_{1}, r^{}_{1}+r^{}_{2}}(\tau) \; .
\end{array}
\label{eq:etaklein}
%     (2.14)
\end{eqnarray}
More information about the properties of the Kein form ${\mathfrak k}^{}_{r^{}_1, r^{}_2}(\tau)$ can be found in Refs.~\cite{ Ding:2019xna}.
%\vspace*{-1cm}
%\vspace{2cm}
\section{The Kronecker product rules of $A^{\prime}_{5}$}
\label{prod_rules}
Here we present only those product rules \cite{Wang:2020lxk} which are relevant to the present model.
\renewcommand\arraystretch{1.5}
%\renewcommand\arraystretch{1.0}
%\vfill
\noindent
\begin{small}
	\begin{tabular}{L{8cm}L{8cm}}
		${\bf 3}^{\prime}_{} \otimes {\bf 3}^{\prime}_{} = {\bf 1}^{}_{\rm s} \oplus {\bf 3}^{\prime}_{\rm a} \oplus {\bf 5}^{}_{\rm s}$ & $\mathbf{3} \otimes \boldsymbol{3}=\mathbf{1}^{}_{\rm s} \oplus \boldsymbol{3}^{}_{\rm a} \oplus \boldsymbol{5}^{}_{\rm s}$   \\ 
$\left\{\begin{array}{l}
		{\bf 1}^{}_{\rm s} : \dfrac{1}{\sqrt3} \left[\alpha^{}_1 \beta^{}_1 + \alpha^{}_2 \beta^{}_3 + \alpha^{}_3 \beta^{}_2\right]\\
		{\bf 3}^{\prime}_{\rm a} : \dfrac{1}{\sqrt2} \left[\begin{array}{c}
		\alpha^{}_{2} \beta^{}_{3}-\alpha^{}_{3} \beta^{}_{2} \\
		\alpha^{}_{1} \beta^{}_{2}-\alpha^{}_{2} \beta^{}_{1} \\
		\alpha^{}_{3} \beta^{}_{1}-\alpha^{}_{1} \beta^{}_{3}
		\end{array}\right] \\
		{\bf 5}^{}_{\rm s} : \dfrac{1}{\sqrt6} \left[\begin{array}{c}
		2 \alpha^{}_{1} \beta^{}_{1}-\alpha^{}_{2} \beta^{}_{3}-\alpha^{}_{3} \beta^{}_{2} \\
		\sqrt{6} \alpha^{}_{3} \beta^{}_{3} \\
		-\sqrt{3} \left(\alpha^{}_{1} \beta^{}_{2}+ \alpha^{}_{2} \beta^{}_{1} \right) \\
		-\sqrt{3} \left( \alpha^{}_{1} \beta^{}_{3}+ \alpha^{}_{3} \beta^{}_{1} \right) \\
		\sqrt{6} \alpha^{}_{2} \beta^{}_{2}
		\end{array}\right]
		\end{array}\right\}.$
		\hspace{4mm}		
		
		& 	$\left\{ \begin{array}{l}
		\mathbf{1}^{}_{\rm s} : \dfrac{1}{\sqrt3} [\alpha^{}_{1} \beta^{}_{1}+\alpha^{}_{2} \beta^{}_{3}+\alpha^{}_{3} \beta^{}_{2}]\\
		{\bf 3}^{}_{\rm a} : \dfrac{1}{\sqrt2} \left[\begin{array}{c}
		\alpha^{}_{2} \beta^{}_{3}-\alpha^{}_{3} \beta^{}_{2} \\
		\alpha^{}_{1} \beta^{}_{2}-\alpha^{}_{2} \beta^{}_{1} \\
		\alpha^{}_{3} \beta^{}_{1}-\alpha^{}_{1} \beta^{}_{3}
		\end{array}\right] \\
		{\bf 5}^{}_{\rm s} : \dfrac{1}{\sqrt6} \left[\begin{array}{c}
		2 \alpha^{}_{1} \beta^{}_{1}-\alpha^{}_{2} \beta^{}_{3}-\alpha^{}_{3} \beta^{}_{2} \\
		-\sqrt{3} \alpha^{}_{1} \beta^{}_{2}-\sqrt{3} \alpha^{}_{2} \beta^{}_{1} \\
		\sqrt{6} \alpha^{}_{2} \beta^{}_{2} \\
		\sqrt{6} \alpha^{}_{3} \beta^{}_{3} \\
		-\sqrt{3} \left(\alpha^{}_{1} \beta^{}_{3}+ \alpha^{}_{3} \beta^{}_{1} \right)
		\end{array}\right]
		\end{array}\right\}.$
		
	\vspace{0.5cm}
	\end{tabular}
	\end{small}
\newpage
\renewcommand\arraystretch{1.5}
	\begin{tabular}{L{7.5cm}L{7.5cm}}
		$ {\bf 3} \otimes {\bf 3}^{\prime}_{} = {\bf 4} \oplus {\bf 5}$ & ${\bf 4} \otimes {\bf 4} = {\bf 1}^{}_{\rm s} \oplus {\bf 3}^{}_{\rm a} \oplus {\bf 3}^{\prime}_{\rm a} \oplus {\bf 4}^{}_{\rm s} \oplus {\bf 5}^{}_{\rm s}$ \\
		 $\left\{\begin{array}{l}
		{\bf 4} : \dfrac{1}{\sqrt3} \left[\begin{array}{c}
		\sqrt{2} \alpha^{}_{2} \beta^{}_{1}+\alpha^{}_{3} \beta^{}_{2} \\
		-\sqrt{2} \alpha^{}_{1} \beta^{}_{2}-\alpha^{}_{3} \beta^{}_{3} \\
		-\sqrt{2} \alpha^{}_{1} \beta^{}_{3}-\alpha^{}_{2} \beta^{}_{2} \\
		\sqrt{2} \alpha^{}_{3} \beta^{}_{1}+\alpha^{}_{2} \beta^{}_{3}
		\end{array}\right] \\
		{\bf 5} : \dfrac{1}{\sqrt3} \left[\begin{array}{c}
		\sqrt{3} \alpha^{}_{1} \beta^{}_{1} \\
		\alpha^{}_{2} \beta^{}_{1}-\sqrt{2} \alpha^{}_{3} \beta^{}_{2} \\
		\alpha^{}_{1} \beta^{}_{2}-\sqrt{2} \alpha^{}_{3} \beta^{}_{3} \\
		\alpha^{}_{1} \beta^{}_{3}-\sqrt{2} \alpha^{}_{2} \beta^{}_{2} \\
		\alpha^{}_{3} \beta^{}_{1}-\sqrt{2} \alpha^{}_{2} \beta^{}_{3}
		\end{array}\right]
		\end{array}\right\}.$
		%\hspace{4cm}

		& $\left\{\begin{array}{l}
		{\bf 1}^{}_{\rm s}: \dfrac{1}{2} \left[\alpha^{}_{1} \beta^{}_{4}+\alpha^{}_{2} \beta^{}_{3}+\alpha^{}_{3} \beta^{}_{2}+\alpha^{}_{4} \beta^{}_{1}\right] \vspace{1cm}\\
		{\bf 3}^{}_{\rm a} : \dfrac{1}{2} \left[\begin{array}{c}
		-\alpha^{}_{1} \beta^{}_{4}+\alpha^{}_{2} \beta^{}_{3}-\alpha^{}_{3} \beta^{}_{2}+\alpha^{}_{4} \beta^{}_{1} \\
		\sqrt{2} \left( \alpha^{}_{2} \beta^{}_{4}- \alpha^{}_{4} \beta^{}_{2} \right) \\
		\sqrt{2} \left( \alpha^{}_{1} \beta^{}_{3}- \alpha^{}_{3} \beta^{}_{1} \right)
		\end{array}\right] \\
		{\bf 3}^{\prime}_{\rm a} : \dfrac{1}{2} \left[\begin{array}{c}
		\alpha^{}_{1} \beta^{}_{4}+\alpha^{}_{2} \beta^{}_{3}-\alpha^{}_{3} \beta^{}_{2}-\alpha^{}_{4} \beta^{}_{1} \\
		\sqrt{2} \left( \alpha^{}_{3} \beta^{}_{4}- \alpha^{}_{4} \beta^{}_{3} \right) \\
		\sqrt{2} \left( \alpha^{}_{1} \beta^{}_{2}- \alpha^{}_{2} \beta^{}_{1} \right)
		\end{array}\right] \\
		{\bf 4}^{}_{\rm s} : \dfrac{1}{\sqrt3} \left[\begin{array}{c}
		\alpha^{}_{2} \beta^{}_{4}+\alpha^{}_{3} \beta^{}_{3}+\alpha^{}_{4} \beta^{}_{2} \\
		\alpha^{}_{1} \beta^{}_{1}+\alpha^{}_{3} \beta^{}_{4}+\alpha^{}_{4} \beta^{}_{3} \\
		\alpha^{}_{1} \beta^{}_{2}+\alpha^{}_{2} \beta^{}_{1}+\alpha^{}_{4} \beta^{}_{4} \\
		\alpha^{}_{1} \beta^{}_{3}+\alpha^{}_{2} \beta^{}_{2}+\alpha^{}_{3} \beta^{}_{1}
		\end{array}\right]\\
		{\bf 5}^{}_{\rm s} : \dfrac{1}{2\sqrt3}\left[\begin{array}{c}
		\sqrt{3} \left( \alpha^{}_{1} \beta^{}_{4}- \alpha^{}_{2} \beta^{}_{3}- \alpha^{}_{3} \beta^{}_{2}+ \alpha^{}_{4} \beta^{}_{1}  \right) \\
		-\sqrt{2} \left(\alpha^{}_{2} \beta^{}_{4}- 2  \alpha^{}_{3} \beta^{}_{3}+ \alpha^{}_{4} \beta^{}_{2} \right) \\
		- \sqrt{2} \left( 2\alpha^{}_{1} \beta^{}_{1}- \alpha^{}_{3} \beta^{}_{4}- \alpha^{}_{4} \beta^{}_{3} \right) \\
		\sqrt{2} \left( \alpha^{}_{1} \beta^{}_{2}+ \alpha^{}_{2} \beta^{}_{1}-2 \alpha^{}_{4} \beta^{}_{4} \right)  \\
		-\sqrt{2} \left( \alpha^{}_{1} \beta^{}_{3} - 2 \alpha^{}_{2} \beta^{}_{2}+ \alpha^{}_{3} \beta^{}_{1}\right)
		\end{array}\right]
		\end{array}\right\}.$
		\end{tabular}\\
		\newpage
\renewcommand\arraystretch{1.4}
	\begin{tabular}{L{6.5cm}}
		${\bf 5} \otimes {\bf 5} = {\bf 1}^{}_{\rm s} \oplus {\bf 3}^{}_{\rm a} \oplus {\bf 3}^{\prime}_{\rm a} \oplus {\bf 4}^{}_{\rm s} \oplus {\bf 4}^{}_{\rm a} \oplus {\bf 5}^{}_{\rm s,1} \oplus {\bf 5}^{}_{\rm s,2} $\vspace{1cm}\\
		$\left\{\begin{array}{l}
		{\bf 1}^{}_{\rm s} : \dfrac{1}{\sqrt5} \left[\alpha^{}_{1} \beta^{}_{1}+\alpha^{}_{2} \beta^{}_{5}+\alpha^{}_{3} \beta^{}_{4}+\alpha^{}_{4} \beta^{}_{3}+\alpha^{}_{5} \beta^{}_{2}\right] \\
		{\bf 3}^{}_{\rm a} : \dfrac{1}{\sqrt10} \left[\begin{array}{c}
		\alpha^{}_{2} \beta^{}_{5}+2 \alpha^{}_{3} \beta^{}_{4}-2 \alpha^{}_{4} \beta^{}_{3}-\alpha^{}_{5} \beta^{}_{2} \\
		-\sqrt{3} \alpha^{}_{1} \beta^{}_{2}+\sqrt{3} \alpha^{}_{2} \beta^{}_{1}+\sqrt{2} \alpha^{}_{3} \beta^{}_{5}-\sqrt{2} \alpha^{}_{5} \beta^{}_{3} \\
		\sqrt{3} \alpha^{}_{1} \beta^{}_{5}+\sqrt{2} \alpha^{}_{2} \beta^{}_{4}-\sqrt{2} \alpha^{}_{4} \beta^{}_{2}-\sqrt{3} \alpha^{}_{5} \beta^{}_{1}
		\end{array}\right]  \\
		{\bf 3}^{\prime}_{\rm a} : \dfrac{1}{\sqrt10} \left[\begin{array}{c}
		2 \alpha^{}_{2} \beta^{}_{5}-\alpha^{}_{3} \beta^{}_{4}+\alpha^{}_{4} \beta^{}_{3}-2 \alpha^{}_{5} \beta^{}_{2} \\
		\sqrt{3} \alpha^{}_{1} \beta^{}_{3}-\sqrt{3} \alpha^{}_{3} \beta^{}_{1}+\sqrt{2} \alpha^{}_{4} \beta^{}_{5}-\sqrt{2} \alpha^{}_{5} \beta^{}_{4} \\
		-\sqrt{3} \alpha^{}_{1} \beta^{}_{4}+\sqrt{2} \alpha^{}_{2} \beta^{}_{3}-\sqrt{2} \alpha^{}_{3} \beta^{}_{2}+\sqrt{3} \alpha^{}_{4} \beta^{}_{1}
		\end{array}\right] \\
		{\bf 4}^{}_{\rm s} : \dfrac{1}{\sqrt30} \left[\begin{array}{l}
		\sqrt{6} \alpha^{}_{1} \beta^{}_{2}+\sqrt{6} \alpha^{}_{2} \beta^{}_{1}- \alpha^{}_{3} \beta^{}_{5}+4  \alpha^{}_{4} \beta^{}_{4}- \alpha^{}_{5} \beta^{}_{3} \\
		\sqrt{6} \alpha^{}_{1} \beta^{}_{3}+4  \alpha^{}_{2} \beta^{}_{2}+\sqrt{6} \alpha^{}_{3} \beta^{}_{1}- \alpha^{}_{4} \beta^{}_{5}- \alpha^{}_{5} \beta^{}_{4} \\
		\sqrt{6} \alpha^{}_{1} \beta^{}_{4}- \alpha^{}_{2} \beta^{}_{3}- \alpha^{}_{3} \beta^{}_{2}+\sqrt{6} \alpha^{}_{4} \beta^{}_{1}+4  \alpha^{}_{5} \beta^{}_{5} \\
		\sqrt{6} \alpha^{}_{1} \beta^{}_{5}- \alpha^{}_{2} \beta^{}_{4}+4  \alpha^{}_{3} \beta^{}_{3}- \alpha^{}_{4} \beta^{}_{2}+\sqrt{6} \alpha^{}_{5} \beta^{}_{1}
		\end{array}\right]\\
		{\bf 4}^{}_{\rm a} : \dfrac{1}{\sqrt10} \left[\begin{array}{c}
		\sqrt{2} \alpha^{}_{1} \beta^{}_{2}-\sqrt{2} \alpha^{}_{2} \beta^{}_{1}+\sqrt{3} \alpha^{}_{3} \beta^{}_{5}-\sqrt{3} \alpha^{}_{5} \beta^{}_{3} \\
		-\sqrt{2} \alpha^{}_{1} \beta^{}_{3}+\sqrt{2} \alpha^{}_{3} \beta^{}_{1}+\sqrt{3} \alpha^{}_{4} \beta^{}_{5}-\sqrt{3} \alpha^{}_{5} \beta^{}_{4} \\
		-\sqrt{2} \alpha^{}_{1} \beta^{}_{4}-\sqrt{3} \alpha^{}_{2} \beta^{}_{3}+\sqrt{3} \alpha^{}_{3} \beta^{}_{2}+\sqrt{2} \alpha^{}_{4} \beta^{}_{1} \\
		\sqrt{2} \alpha^{}_{1} \beta^{}_{5}-\sqrt{3} \alpha^{}_{2} \beta^{}_{4}+\sqrt{3} \alpha^{}_{4} \beta^{}_{2}-\sqrt{2} \alpha^{}_{5} \beta^{}_{1}
		\end{array}\right] \\
		{\bf 5}^{}_{\rm s,1} : \dfrac{1}{\sqrt14} \left[\begin{array}{c}
		2 \alpha^{}_{1} \beta^{}_{1}+\alpha^{}_{2} \beta^{}_{5}-2 \alpha^{}_{3} \beta^{}_{4}-2 \alpha^{}_{4} \beta^{}_{3}+\alpha^{}_{5} \beta^{}_{2} \\
		\alpha^{}_{1} \beta^{}_{2}+\alpha^{}_{2} \beta^{}_{1}+\sqrt{6} \alpha^{}_{3} \beta^{}_{5}+\sqrt{6} \alpha^{}_{5} \beta^{}_{3} \\
		-2 \alpha^{}_{1} \beta^{}_{3}+\sqrt{6} \alpha^{}_{2} \beta^{}_{2}-2 \alpha^{}_{3} \beta^{}_{1} \\
		-2 \alpha^{}_{1} \beta^{}_{4}-2 \alpha^{}_{4} \beta^{}_{1}+\sqrt{6} \alpha^{}_{5} \beta^{}_{5} \\
		\alpha^{}_{1} \beta^{}_{5}+\sqrt{6} \alpha^{}_{2} \beta^{}_{4}+\sqrt{6} \alpha^{}_{4} \beta^{}_{2}+\alpha^{}_{5} \beta^{}_{1}
		\end{array}\right]   \\ 
		{\bf 5}^{}_{\rm s,2} : \dfrac{1}{\sqrt14} \left[\begin{array}{c}
		2 \alpha_{1} \beta_{1}-2 \alpha_{2} \beta_{5}+\alpha_{3} \beta_{4}+\alpha_{4} \beta_{3}-2 \alpha_{5} \beta_{2} \\
		-2 \alpha_{1} \beta_{2}-2 \alpha_{2} \beta_{1}+\sqrt{6} \alpha_{4} \beta_{4} \\
		\alpha_{1} \beta_{3}+\alpha_{3} \beta_{1}+\sqrt{6} \alpha_{4} \beta_{5}+\sqrt{6} \alpha_{5} \beta_{4} \\
		\alpha_{1} \beta_{4}+\sqrt{6} \alpha_{2} \beta_{3}+\sqrt{6} \alpha_{3} \beta_{2}+\alpha_{4} \beta_{1} \\
		-2 \alpha_{1} \beta_{5}+\sqrt{6} \alpha_{3} \beta_{3}-2 \alpha_{5} \beta_{1}
		\end{array}\right]
		\end{array}\right\}.$
		\vspace{0.6cm}
	\end{tabular}\\
	%\newpage
\section{Higher Order Yukawa couplings}
\label{HOYC}
All higher order Yukawa couplings are expressed in terms of the elements of $Y^{(1)}_{\widehat{\bm 6}}$ Yukawa coupling expressed as
%\vspace{-2cm}
\begin{eqnarray}
Y^{(1)}_{\widehat{\bf 6}} =
\left[\begin{matrix}
Y^{}_1 \\
Y^{}_2 \\
Y^{}_3 \\
Y^{}_4 \\
Y^{}_5 \\
Y^{}_6 \\
\end{matrix}\right]
=\left[\begin{matrix}
\widehat{e}^{}_{1}-3 \, \widehat{e}^{}_6 \\
5\sqrt{2} \, \widehat{e}^{}_2 \\
10 \, \widehat{e}^{}_3 \\
10 \, \widehat{e}^{}_4 \\
5\sqrt{2} \, \widehat{e}^{}_5 \\
-3 \, \widehat{e}^{}_1-\widehat{e}^{}_6 \\
\end{matrix}\right] \; .
\label{eq:Y_16}
%     (2.21)
\end{eqnarray}
The Yukawa couplings used in our model are expressed below and the other couplings seen in the tensor product are expressed in \cite{Wang:2020lxk}
\begin{eqnarray}
Y^{(2)}_{{\bf 3}} &=& \left[Y^{(1)}_{\widehat{\bf 6}} \otimes Y^{(1)}_{\widehat{\bf 6}}\right]^{}_{{\bf 3}^{}_{{\rm s},1}} =
-3 \left[
\begin{array}{c}
\widehat{e}^{2}_1 - 36\, \widehat{e}_1 \widehat{e}^{}_6 - \widehat{e}_6^2 \\
5 \sqrt{2}\, \widehat{e}_2^{} (\widehat{e}^{}_1 - 3\, \widehat{e}_6^{}) \\
5 \sqrt{2}\, \widehat{e}_5^{} (3\, \widehat{e}^{}_1 + \widehat{e}_6^{}) \\
\end{array}
\right]
= -3 \left[
\begin{array}{c}
Y^2_1-3Y^{}_1Y^{}_6-Y^2_{6} \\
Y^{}_1 Y^{}_2 \\
-Y^{}_5 Y^{}_6 \\
\end{array}\right], \nonumber \\
\end{eqnarray}
\begin{eqnarray}
Y^{(4)}_{\bf 3} &=&  \left[Y^{(1)}_{\widehat{\bf 6}} \otimes Y^{(3)}_{\widehat{\bf 6},2}\right]^{}_{{\bf 3}^{}_{{\rm s},1}} = \frac{\sqrt{3}}{4}
\left[
\begin{array}{c}
\left(Y_1^2+Y_6^2\right)\left(7 Y_1^2-18  Y_1^{} Y_6^{}-7 Y_6^2\right) \\
Y_2^{} \left(13 Y_1^3-3 Y_1^{2} Y_6-29 Y_1^{} Y_6^2 -9
Y_6^3\right) \\
-Y_5^{} \left(9 Y_1^3-29 Y_1^2 Y_6^{} +3 Y_1^{} Y_6^2 +13
Y_6^3\right) \\
\end{array}
\right]  , \nonumber \\
\end{eqnarray}
\begin{eqnarray}
Y^{(6)}_{{\bf 3},1} &=& \left[ Y^{(1)}_{\widehat{\bf 6}} \otimes Y^{(5)}_{\widehat{\bf 2}^{\prime}_{}}\right]^{}_{\bf 3} =\dfrac{9\sqrt{2}}{16} \left(Y^2_1- 4 Y^{}_1 Y^{}_6 -Y^2_6\right) \left[
\begin{array}{c}
(Y_1-3 Y_6) (3 Y_1+Y_6) \left(3 Y_1^2-2 Y_1
Y_6-3 Y_6^2\right) \\
2 Y_2^{} \left(2 Y_1^3-9 Y_1 Y_6^2-3 Y_6^3\right) \\
2 Y_5^{} \left(3 Y_1^3-9 Y_1^2 Y_6+2 Y_6^3\right) \\
\end{array}
\right] \; , \nonumber\\
\label{eq:Y6_1}
\end{eqnarray}
\begin{eqnarray}
Y^{(6)}_{{\bf 3},2} &=& \left[ Y^{(1)}_{\widehat{\bf 6}} \otimes Y^{(5)}_{\widehat{\bf 6},1}\right]^{}_{{\bf 3}^{}_{{\rm s},1}} =3\sqrt{2} \left(Y_1^4 - 3 Y_1^3 Y_6^{} - Y_1^2 Y_6^2 + 3 Y_1^{} Y_6^3 + Y_6^4\right) \left[
\begin{array}{c}
Y_1^2-3 Y_1^{} Y_6^{}-Y_6^2 \\
Y_1^{} Y_2^{} \\
-Y_5^{} Y_6^{} \\
\end{array}
\right] \; , \nonumber \\
\end{eqnarray}
\begin{eqnarray}
Y^{(6)}_{{\bf 4},1} &=& \left[ Y^{(1)}_{\widehat{\bf 6}} \otimes Y^{(5)}_{\widehat{\bf 2}}\right]^{}_{\bf 4} =-\dfrac{3}{4} \left(Y^2_1- 4 Y^{}_1 Y^{}_6 -Y^2_6\right)^2_{} \left[
\begin{array}{c}
-\sqrt{2} Y_2^{} (3 Y_1+Y_6) \\
Y_3^{} (Y_1+Y_6) \\
Y_4^{} (Y_1-Y_6) \\
\sqrt{2} Y_5^{} (Y_1-3 Y_6) \\
\end{array}
\right] \; , \nonumber \\
\end{eqnarray}
\begin{eqnarray}
Y^{(6)}_{{\bf 4},2} &=& \left[ Y^{(1)}_{\widehat{\bf 6}} \otimes Y^{(5)}_{\widehat{\bf 2}^{\prime}_{}}\right]^{}_{\bf 4} =-\dfrac{\sqrt{6}}{8} \left(Y^2_1- 4 Y^{}_1 Y^{}_6 -Y^2_6\right) \left[
\begin{array}{c}
\sqrt{2} Y_2^{} \left(Y_1^3+11 Y_1^2 Y_6^{}+19 Y_1^{}
Y_6^2+5 Y_6^3\right) \\
Y_3^{} \left(13 Y_1^3-31 Y_1^2 Y_6^{}-17 Y_1^{}
Y_6^2-Y_6^3\right) \\
Y_4^{} \left(Y_1^3-17 Y_1^2 Y_6^{}+31 Y_1^{} Y_6^2+13
Y_6^3\right) \\
\sqrt{2} Y_5^{} \left(5 Y_1^3-19 Y_1^2 Y_6^{} + 11 Y_1^{}
Y_6^2-Y_6^3\right) \\
\end{array}
\right]\; , \nonumber \\
\end{eqnarray}
\begin{eqnarray}
Y^{(6)}_{{\bf 5},1} &=& \left[ Y^{(1)}_{\widehat{\bf 6}} \otimes Y^{(5)}_{\widehat{\bf 4}}\right]^{}_{{\bf 5},2} =\dfrac{\sqrt{10}}{8} \left(Y^2_1- 4 Y^{}_1 Y^{}_6 -Y^2_6\right) \left[
\begin{array}{c}
\sqrt{3} (Y_1^{}-3 Y_6^{}) (3 Y_1^{}+Y_6^{})
\left(Y_1^2+Y_6^2\right) \\
-2 Y_2^{} (2 Y_1^{}+Y_6^{}) \left(2 Y_1^2-3 Y_1^{}
Y_6^{}-Y_6^2\right) \\
\sqrt{2} Y_3^{} \left(Y_1^3+2 Y_1^2 Y_6^{}-11 Y_1^{}
Y_6^2-4 Y_6^3\right) \\
\sqrt{2} Y_4^{} \left(4 Y_1^3-11 Y_1^2 Y_6^{}-2 Y_1^{}
Y_6^2+Y_6^3\right) \\
2 Y_5^{} (Y_1^{}-2 Y_6^{}) \left(Y_1^2-3 Y_1^{} Y_6^{}-2
Y_6^2\right) \\
\end{array}
\right] \; , \nonumber \\
\end{eqnarray}
\begin{eqnarray}
Y^{(6)}_{{\bf 5},2} &=& \left[ Y^{(1)}_{\widehat{\bf 6}} \otimes Y^{(5)}_{\widehat{\bf 6},1}\right]^{}_{{\bf 5}^{}_{\rm s}} =-\dfrac{1}{\sqrt{2}}\left(Y_1^4 - 3 Y_1^3 Y_6^{} - Y_1^2 Y_6^2 + 3 Y_1^{} Y_6^3 + Y_6^4\right) \left[
\begin{array}{c}
\sqrt{2} \left(Y_1^2+Y_6^2\right) \\
2 \sqrt{6} Y_2^{} (2 Y_1^{}+Y_6^{}) \\
-\sqrt{3} Y_3^{} (3 Y_1^{}-Y_6^{}) \\
\sqrt{3} Y_4^{} (Y_1^{}+3 Y_6^{}) \\
-2 \sqrt{6} Y_5^{} (Y_1^{}-2 Y_6^{}) \\
\end{array}
\right] \; . \nonumber\\
\end{eqnarray}

\bibliographystyle{my-JHEP}
\bibliography{a5_inverse}

\end{document}